\documentclass[prd,superscriptaddress,nofootinbib,onecolumn]{revtex4}
\usepackage{amsmath,amssymb,epsfig}

\usepackage{graphicx}
\usepackage{bm}

\usepackage{scalerel}
\usepackage{tikz}
\usetikzlibrary{svg.path}

\definecolor{orcidlogocol}{HTML}{A6CE39}
\tikzset{
	orcidlogo/.pic={
		\fill[orcidlogocol] svg{M256,128c0,70.7-57.3,128-128,128C57.3,256,0,198.7,0,128C0,57.3,57.3,0,128,0C198.7,0,256,57.3,256,128z};
		\fill[white] svg{M86.3,186.2H70.9V79.1h15.4v48.4V186.2z}
		svg{M108.9,79.1h41.6c39.6,0,57,28.3,57,53.6c0,27.5-21.5,53.6-56.8,53.6h-41.8V79.1z M124.3,172.4h24.5c34.9,0,42.9-26.5,42.9-39.7c0-21.5-13.7-39.7-43.7-39.7h-23.7V172.4z}
		svg{M88.7,56.8c0,5.5-4.5,10.1-10.1,10.1c-5.6,0-10.1-4.6-10.1-10.1c0-5.6,4.5-10.1,10.1-10.1C84.2,46.7,88.7,51.3,88.7,56.8z};
	}
}

\newcommand\orcidicon[1]{\href{https://orcid.org/#1}{\mbox{\scalerel*{
				\begin{tikzpicture}[yscale=-1,transform shape]
				\pic{orcidlogo};
				\end{tikzpicture}
			}{1}}}}

\usepackage{natbib,ifthen}
\usepackage{xcolor}
\usepackage[hidelinks]{hyperref}

\begin{document}

\newcommand{\fixme}[1]{{\textbf{Fixme: #1}}}
\newcommand{\detD}{{\det\!\cld}}
\newcommand{\clh}{\mathcal{H}}
\newcommand{\ud}{{\rm d}}
\renewcommand{\eprint}[1]{\href{http://arxiv.org/abs/#1}{#1}}
\newcommand{\adsurl}[1]{\href{#1}{ADS}}
\newcommand{\ISBN}[1]{\href{http://cosmologist.info/ISBN/#1}{ISBN: #1}}
\newcommand{\jcap}{J.\ Cosmol.\ Astropart.\ Phys.}
\newcommand{\mnras}{Mon.\ Not.\ R.\ Astron.\ Soc.}
\newcommand{\progress}{Rep.\ Prog.\ Phys.}
\newcommand{\prlett}{Phys.\ Rev.\ Lett.}
\newcommand{\procspie}{Proc.\ SPIE}
\newcommand{\na}{New Astronomy}
\newcommand{\apjl}{ApJ.\ Lett.}
\newcommand{\physrep}{Physics Reports}
\newcommand{\aap}{A\&A}
\newcommand{\aapr}{A\&A Rev.}
\newcommand{\araa}{ARA\&A}
\newcommand{\pasj}{PASJ}
\newcommand{\aaps}{A\&AS}

\newcommand{\ThreeJSymbol}[6]{\begin{pmatrix}
#1 & #3 & #5 \\
#2 & #4 & #6
 \end{pmatrix}}

\newcommand{\mathbfss}[1]{\mathbf{#1}}


\title[Angular power spectrum likelihoods]{The non-Gaussian likelihood of weak lensing power spectra}


\author{Alex Hall \orcidicon{0000-0002-3139-8651}}
\email{ahall@roe.ac.uk}
\author{Andy Taylor}
\affiliation{Institute for Astronomy, University of Edinburgh, Royal Observatory, Blackford Hill, Edinburgh, EH9 3HJ, U.K.}



\begin{abstract}
The power spectrum of weak lensing fluctuations has a non-Gaussian distribution due to its quadratic nature. On small scales the Central Limit Theorem acts to Gaussianize this distribution but non-Gaussianity in the signal due to gravitational collapse is increasing and the functional form of the likelihood is unclear. Analyses have traditionally assumed a Gaussian likelihood with non-linearity incorporated into the covariance matrix; here we provide the theory underpinning this assumption. We calculate, for the first time, the leading-order correction to the distribution of angular power spectra from non-Gaussianity in the underlying signal and study the transition to Gaussianity. Our expressions are valid for an arbitrary number of correlated maps and correct the Wishart distribution in the presence of weak (but otherwise arbitrary) non-Gaussianity in the signal. Surprisingly, the resulting distribution is not equivalent to an Edgeworth expansion. The leading-order effect is to broaden the covariance matrix by the usual trispectrum term, with residual skewness sourced by the trispectrum and the square of the bispectrum. Using lognormal lensing maps we demonstrate that our likelihood is uniquely able to model both large and mildly non-linear scales. We provide easy-to-compute statistics to quantify the size of the non-Gaussian corrections. We show that the full non-Gaussian likelihood can be accurately modelled as a Gaussian on small, non-linear scales. On large angular scales non-linearity in the lensing signal imparts a negligible correction to the likelihood, which takes the Wishart form in the full-sky case. Our formalism is equally applicable to any kind of projected field.
\end{abstract}

\maketitle

\section{Introduction}
\label{sec:intro}

In the coming decade several large astronomy projects aim to place percent-level constraints on cosmic acceleration, testing the $\Lambda$CDM cosmological model and its extensions. These include the European Space Agency's Euclid mission\footnote{\url{https://sci.esa.int/web/euclid}}, the Vera C. Rubin Observatory\footnote{\url{https://www.lsst.org/about}}, and the Nancy Grace Roman Space Telescope\footnote{\url{https://roman.gsfc.nasa.gov/}}. These `Stage-IV' facilities will each image roughly a billion galaxies, with redshifts derived from photometry calibrated against deep spectroscopic training samples.

Upcoming surveys will use these galaxies to measure the weak gravitational lensing signal (see Refs.~\cite{2015RPPh...78h6901K, 2018ARA&A..56..393M} for recent reviews), which directly probes the underlying matter distribution and provides the opportunity to derive cosmological constraints from its statistical properties. Current lensing surveys are now achieving precision on certain $\Lambda$CDM cosmological parameter combinations at the level of a few percent, and providing useful constraints on the properties of dark energy~\cite{2020PASJ...72...16H, 2021A&A...646A.140H, 2021arXiv210513549D}. The influence of weak lensing in constraining cosmological parameters is likely to increase with the advent of Stage-IV surveys, if systematic errors can be kept under control.

It is now fairly common practice to combine measurements of the cosmic shear signal with maps of the angular positions of galaxies, sometimes referred to as `3$\times$2 pt' analysis~\cite{2018PhRvD..98d3526A}. The relatively coarse precision of photometric redshifts means that binning galaxies into broad redshift bins is reasonably lossless, a technique called `tomography'~\cite{1999ApJ...522L..21H, 2010A&A...516A..63S, 2013MNRAS.432.2433H, 2016ApJ...824...77J}. Combining tomographic galaxy clustering and weak lensing are examples of the use of \emph{projected} maps of dark matter (or discrete tracers thereof) to constrain cosmological models. 

The baseline parameter constraints from current and upcoming surveys using projected fields will be derived from two-point statistics, such as correlation functions or power spectra, combined with a likelihood function. Accurate cosmological inference using this approach requires the likelihood to be specified accurately. The functional form of the likelihood has traditionally been assumed to be Gaussian in the measured summary statistics, the rationale being that the Central Limit Theorem (CLT) drives the distribution of these statistics towards Gaussianity. However, it has long been known that the distribution of weak lensing or galaxy clustering two-point statistics is in fact \emph{not} Gaussian, for several reasons. Firstly, two-point statistics are non-linear (quadratic) functions of the data. If the data are Gaussian distributed, power spectra and correlation functions are expected to be non-Gaussian as a result, which can easily be seen by noting that the variance of the field (the correlation function at zero lag) cannot be negative. On the full sky the distribution of power spectra of Gaussian fields can be written down analytically (e.g.~Ref.~\cite{2006MNRAS.372.1104P}), but in the presence of a survey mask (required to mitigate foreground contamination) the naive estimator (i.e.~the pseudo-$C_l$ estimator) has a distribution that is complicated to compute and must typically be approximated~\cite{2008PhRvD..77j3013H, 2020MNRAS.491.3165U}.

A second source of non-Gaussianity in the likelihood arises from the fact that the fields themselves are not Gaussian due to non-linear gravitational collapse. This issue is inextricably tied to the efficacy of the CLT to Gaussianize the estimator, because both the number of degrees of freedom being compressed by the estimator (e.g.~the number of galaxy pairs of a given angular separation) and the strength of the signal non-linearity increase as the minimum angular scale is lowered. Non-Gaussianity of the signal (the assumption of Gaussian noise is typically safe) is usually incorporated by modifying the covariance matrix of the two-point statistic with a term derived from the connected four-point function (trispectrum) of the signal, while keeping the Gaussian functional form (e.g.~\cite{2019PASJ...71...43H, 2021JCAP...03..067N, 2021arXiv211006947L}).

Additionally, non-Gaussianity in the likelihood can also arise due to the presence of non-Gaussian systematic residuals, or when the covariance has been estimated from simulations~\cite{2016MNRAS.456L.132S} or combined with an analytic model~\cite{2019MNRAS.483..189H}.

Recognising that the approximation of Gaussianity may contribute bias or imprecision to parameter constraints, several works have successfully measured non-Gaussianity in the distribution of two-point statistics from simulations. Ref.~\cite{2000ApJ...544..597S} was an early work to find evidence for non-Gaussian moments of power spectrum estimates, but Ref.~\cite{2009A&A...504..689H} and Ref.~\cite{2009A&A...504..705S} represent the first attempts to incorporate non-Gaussianity into the likelihood of a weak lensing survey. The approach taken in Ref.~\cite{2009A&A...504..689H} was to model the data vector (correlation functions in their case) as the sum of independent non-Gaussian components, using the \texttt{fastICA} Independent Component Analysis method to construct the likelihood. Subsequent analytic work by Ref.~\cite{2011A&A...534A..76K} and  Ref.~\cite{2013A&A...556A..70W} further established the non-Gaussianity of the weak lensing correlation functions. Ref.~\cite{2018MNRAS.473.2355S} introduced diagnostic statistics to identify non-Gaussianity in cosmological likelihood functions, and evidence for non-Gaussianity in a weak lensing likelihood was found in Ref.~\cite{2018MNRAS.477.4879S}. Ref.~\cite{2019MNRAS.485.2956H} studied non-Gaussianity in galaxy clustering power spectra, building on and extending the work of Ref.~\cite{2009A&A...504..689H}, and found non-negligible (albeit sub-$1\sigma$) biases in parameters. In the context of weak lensing correlation functions from Stage-IV surveys, Ref.~\cite{2020MNRAS.499.2977L} identified significant skewness and kurtosis in the shear correlation functions and used a Principal Component Analysis to model the resulting non-Gaussian likelihood, but found that a Gaussian approximation was sufficient for $\Lambda$CDM parameters and an LSST-like survey. This conclusion is supported by the work of Ref.~\cite{2019PhRvD.100b3519T} who used likelihood-free methods to establish the validity of a Gaussian approximation for Stage-IV surveys and $\Lambda$CDM parameters. Recently, Ref.~\cite{2020PhRvD.102j3507D} demonstrated clear non-Gaussianity in the likelihood of weak lensing power spectra, and presented a method to model this using normalizing flows. Ref.~\cite{2021MNRAS.508.3125F} have also studied the impact of non-Gaussian likelihoods on parameter constraints, finding a negligible impact for the Dark Energy Survey data vector.

What these works have established is that while non-Gaussianity is clearly present in the weak lensing likelihood, the impact of the Gaussian assumption on the posterior of parameters appears to be very modest, at least in $\Lambda$CDM models. However, there has been relatively little attempt to distinguish between the non-Gaussianity that should arise on survey scales due to the finite number of modes (that makes a Gamma distribution more accurate than a Gaussian) from the non-Gaussianity that could arise on small scales due to the CLT being inefficient at Gaussianizing the summary statistics in the face of increasing non-linearity in the signal. The use of the power spectrum as the statistic of choice aids in distinguishing between these two effects, as well as allowing the modelling to be simplified due to the diagonal covariance that arises for Gaussian fields. As well as being unable to distinguish between various sources of non-Gaussianity, most of the works modelling non-Gaussian likelihoods have used numerical techniques or ad-hoc approximations to capture the corrections, often with little physical justification. This is partly because the full non-Gaussian distribution of the shear or galaxy clustering signal cannot be written down.

In this work, we incorporate the effects of a non-Gaussian signal into the likelihood of the angular power spectrum, and study how this impacts various posterior distributions of interest. To make analytic progress we assume that the signal non-Gaussianity is sufficiently weak that it can be treated perturbatively. While this is a constraining assumption, it will allow us to study in detail how the various sources of non-Gaussianity interact and make sense of the results of the numerical studies referenced above. Our approach is more in the spirit of Ref.~\cite{2006PhRvD..73b3517S}, who studied similar kinds of effects in the power spectra of the weakly lensed cosmic microwave background, than the numerical works that have sought to model the full non-perturbative likelihood. We argue that an analytic approach building from the well-understood case of Gaussian fields and full sky coverage offers a valuable way of understanding the Gaussianity, or otherwise, of the likelihood function. Having a tractable model that can be quickly computed (and sampled from) also allows the effects of non-Gaussianity to be studied for any cosmological model, rather than being limited to $\Lambda$CDM.

This paper is structured as follows. In Section~\ref{sec:likes} we present the main expression used in this work, Equation~\eqref{eq:fullcorrection}, and study its properties. In Section~\ref{sec:quant} we detail the simulations used to validate our likelihood and quantify the corrections to the posterior distributions. We present our conclusions in Section~\ref{sec:conc}. In a series of appendices we present the derivation of the likelihood function, study the impact of alternative angular binning choices, and test the impact of the source redshift.

\section{The likelihood of angular power spectrum estimates}
\label{sec:likes}

Consider a survey that has made noisy maps $d^i(\hat{\mathbf{n}})$ of the large-scale scale structure across the full sky in $p$ tomographic redshift bins labelled by the index $i$. Taking the spherical harmonic coefficients $a^i_{lm}$ of these maps and arranging them into a vector of length $p$, the angular power spectra including all cross-bin pairs can be packaged into a $p \times p$ matrix $\hat{\mathbfss{C}}_l$ given by
\begin{equation}
    \nu \hat{\mathbfss{C}}_l \equiv \sum_{m=-l}^l \mathbf{a}_{lm} \mathbf{a}_{lm}^\dagger,
    \label{eq:Chatdef}
\end{equation}
where $\nu \equiv 2l + 1$ is the number of angular degrees of freedom at each $l$, and the spherical multipoles of the data consist of signal plus noise as $a^i_{lm} = s^i_{lm} + n^i_{lm}$. We will assume the maps are spin-0 (i.e.~scalar), although this could easily be relaxed; in the case of shear data for example we can define the vector $\mathbf{a}_{lm}$ to include two entries per redshift bin corresponding to the shear $E$ and $B$ modes. Similarly, projected galaxy number counts may be included in the data vector for a full $3\times 2$pt analysis. The hatted power spectrum in Equation~\eqref{eq:Chatdef} denotes a measured, noisy quantity. We will assume that the maps are purely real, and hence all multipoles obey 
\begin{equation}
    \mathbf{a}_{l-m} = (-1)^m \mathbf{a}^*_{lm}.
    \label{eq:reality}
\end{equation}

Our goal is to compute the sampling distribution of these power spectrum estimates, $p(\{\hat{\mathbfss{C}}_l\})$, with braces indicating the full set of angular multipoles $l$, given knowledge of the underlying statistical properties of the noise and the signal. The sampling distribution considered as a function of these properties is the likelihood function. The power spectra are quadratic in the data and so are expected to obey Gaussian statistics only when the Central Limit Theorem is effective. The other source of non-Gaussianity is from the signal itself, which when measured in the late Universe on small scales has higher-order moments due to non-linear structure formation.


As discussed in Appendix~\ref{app:Edge} we cannot proceed by performing an Edgeworth expansion of the power spectrum distribution around its zero-order Wishart form, so instead we develop a new approach and begin by considering the statistics of the underlying data. Assuming the signal and noise are independent, the joint distribution of the power spectra (multiplied by $\nu$) is
\begin{equation}
    p(\{\nu \hat{\mathbfss{C}}_l\}) = \int \mathrm{d}\{\mathbf{s}_{lm}\} \int \mathrm{d}\{\mathbf{n}_{lm}\} \left[\prod_{l=l_{{\rm min}}}^{l_{{\rm max}}} \delta_D\left(\nu \hat{\mathbfss{C}}_l - \sum_m \mathbf{a}_{lm} \mathbf{a}_{lm}^\dagger\right) \right]p_s(\{\mathbf{s}_{lm}\}) p_n(\{\mathbf{n}_{lm}\}),
    \label{eq:pnuCl_1_pne1}
\end{equation}
where the probability density $p(\{\nu \hat{\mathbfss{C}}_l\})$ is defined with a measure over \emph{real and symmetric matrices}, which have $n\equiv p(p+1)/2$ independent elements. We will expand the Dirac delta in terms of real and symmetric $p$-dimensional matrices $\mathbfss{J}_l$ as
\begin{equation}
    \delta_D\left(\nu \hat{\mathbfss{C}}_l - \sum_m \mathbf{a}_{lm} \mathbf{a}_{lm}^\dagger\right) = 2^{\frac{p(p-1)}{2}}\int \frac{\mathrm{d}\mathbfss{J}_{l}}{(2\pi)^n} e^{i \mathrm{Tr}\left[\mathbfss{J}_l \left(\nu \hat{\mathbfss{C}}_l - \sum_m \mathbf{a}_{lm}\mathbf{a}_{lm}^\dagger\right) \right]},
    \label{eq:DiracD_pne1}
\end{equation}
where we integrate over all real symmetric matrices and the volume element is
\begin{equation}
    \mathrm{d}\mathbfss{J}_l = \mathrm{d}J_l^{1,1} \mathrm{d}J_l^{1,2} \dots \mathrm{d}J_l^{1,p} \mathrm{d}J_l^{2,2} \mathrm{d}J_l^{2,3} \dots \mathrm{d}J_l^{p,p}.
\end{equation}
Note that the $2^{\frac{p(p-1)}{2}}$ factor is needed since the off-diagonal elements of $\mathbfss{J}_l$ are double counted in the summation in the exponential, and hence we need to redefine each of the $p(p-1)/2$ off-diagonal terms in the integration with a factor of $2$.

We expand the signal distribution in Fourier modes as
\begin{equation}
    p(\{\mathbf{s}_{lm}\}) = \int \frac{\mathrm{d}\{\mathbf{k}_{lm}\}}{(2\pi)^{pN}} \, \phi_s(\{\mathbf{k}_{lm}\}) \, e^{i\sum_{lm}\mathbf{k}_{lm}^\dagger \mathbf{s}_{lm}},
    \label{eq:FFTslm_pne1}
\end{equation}
where $N \equiv \sum_{l=l_{\mathrm{min}}}^{l_{\mathrm{max}}}2l+1$ is the total number of modes per tomographic bin, and $\phi_s$ is the characteristic function of the signal\footnote{We will define characteristic functions this way throughout, but we caution the reader that other authors sometimes define it with a complex conjugate.}. We can, and will, define the `wave numbers' $\mathbf{k}_{lm}$ to obey reality conditions as in Equation~\eqref{eq:reality}. The corresponding characteristic function of the noise is $\phi_n$. 

Inserting Equations~\eqref{eq:DiracD_pne1} and~\eqref{eq:FFTslm_pne1} into Equation~\eqref{eq:pnuCl_1_pne1}, changing integration variables and doing an integral over $\mathbf{a}_{lm}$ assuming $\mathbfss{J}_l$ is non-singular\footnote{The set of singular $\mathbfss{J}_l$ has zero measure, and as the integrand is finite for this set the contribution from singular modes is vanishingly small in the limit.} gives
\begin{equation}
    p(\{\nu \hat{\mathbfss{C}}_l\}) = 2^{\frac{\lambda p(p-1)}{2}}\int \frac{\mathrm{d}\{ \mathbfss{J}_l \}}{(2\pi)^{n\lambda}} e^{i\sum_l \mathrm{Tr}(\nu \mathbfss{J}_l \hat{\mathbfss{C}}_l)} \left(\prod_{l=l_{\mathrm{min}}}^{l_{\mathrm{max}}} \lvert 2i\mathbfss{J}_l \rvert^{-\frac{\nu}{2}}\right) \int \frac{\mathrm{d}\{\mathbf{k}_{lm}\}}{(2\pi)^{\frac{pN}{2}}} \phi_s(\{\mathbf{k}_{lm}\}) e^{-\frac{1}{2}\sum_{lm}\mathbf{k}_{lm}^\dagger (2i\mathbfss{J}_l)^{-1}\mathbf{k}_{lm}} \phi_n(\{\mathbf{k}_{lm}\}),
\end{equation}
where $\lambda \equiv l_{\mathrm{max}} - l_{\mathrm{min}} + 1$.

Now, we define the characteristic function of $\{\nu \hat{\mathbfss{C}}_l\}$ over its independent elements as
\begin{equation}
    \phi_{\{\nu \hat{\mathbfss{C}}_l\}}(\{\mathbfss{J}_l\}) = \int \mathrm{d}\{ \nu\hat{\mathbfss{C}}_l \} e^{-i\sum_l \mathrm{Tr}(\nu \mathbfss{J}_l \hat{\mathbfss{C}}_l)} p(\{\nu \hat{\mathbfss{C}}_l\}).
\end{equation}
This gives, finally,
\begin{equation}
    \phi_{\{\nu \hat{\mathbfss{C}}_l\}}(\{\mathbfss{J}_l\}) = \left(\prod_{l=l_{\mathrm{min}}}^{l_{\mathrm{max}}} \lvert 2i\mathbfss{J}_l \rvert^{-\frac{\nu}{2}}\right) 
    \int \frac{\mathrm{d}\{\mathbf{k}_{lm}\}}{(2\pi)^{\frac{pN}{2}}} \phi_s(\{\mathbf{k}_{lm}\}) e^{-\frac{1}{2}\sum_{lm}\mathbf{k}_{lm}^\dagger (2i\mathbfss{J}_l)^{-1}\mathbf{k}_{lm}} \phi_n(\{\mathbf{k}_{lm}\}),
    \label{eq:char_1}
\end{equation}
i.e.~a mode coupling integral between the characteristic functions of the signal, the noise, and a Gaussian in the `wave number' of the field $\mathbf{k}_{lm}$ having covariance $2i\mathbfss{J}$, where $\mathbfss{J}$ is the `wave number' of the power spectra. 

\subsection{The known case of a Gaussian signal and Gaussian noise}
\label{subsec:W}

So far, we have not assumed anything about the statistics of the signal or the noise, only that they are independent. In the case of Gaussian signal and noise, the characteristic functions $\phi_s$ and $\phi_n$ also take a Gaussian form, and the integrals in Equation~\eqref{eq:char_1} all decouple. Each integration has a Gaussian form that can be done analytically, with the result being a product of Wishart characteristic functions
\begin{equation}
    \phi_{\{\nu \hat{\mathbfss{C}}_l\}}(\{\mathbfss{J}_l\}) = \prod_{l=l_{\mathrm{min}}}^{l_{\mathrm{max}}} \lvert \mathbfss{I} + 2i\mathbfss{J}_l\mathbfss{C}_l \rvert^{-\frac{\nu}{2}},
    \label{eq:wishchar}
\end{equation}
where $\mathbfss{C}_l$ is the expected total (signal plus noise) covariance matrix of the map. The inverse Fourier transform of this can be computed analytically (see Appendix~\ref{app:pne1}), and gives the well-known distribution of the power spectra as a product of Wishart distributions (e.g., Ref.~\cite{2006MNRAS.372.1104P})
\begin{align}
    p(\{ \hat{\mathbfss{C}}_l\}) &=  \prod_{l=l_{\mathrm{min}}}^{l_{\mathrm{max}}} \frac{ \lvert  \hat{\mathbfss{C}}_l \rvert^{(\nu - p -1)/2}}{2^{\frac{p\nu}{2}}  \Gamma_p\left(\frac{\nu}{2}\right) \left \vert \mathbfss{C}_l /\nu \right \vert^{\frac{\nu}{2}}}e^{-\frac{\nu}{2}\mathrm{Tr}(\mathbfss{C}_l^{-1}\hat{\mathbfss{C}}_l)} \nonumber \\
    &\equiv \prod_{l=l_{\mathrm{min}}}^{l_{\mathrm{max}}} W_p(\hat{\mathbfss{C}}_l; \mathbfss{C}_l/\nu, \nu),
    \label{eq:wishdist}
\end{align}
where $W_p$ is the $p$-dimensional Wishart density and $\Gamma_p$ is the multivariate Gamma function.

Each $\hat{\mathbfss{C}}_l$ estimate thus has a Wishart distribution with scale matrix $\mathbfss{C}_l/\nu$ and degrees of freedom $\nu$ with $\nu > p-1$. As expected, the mean of each $\hat{\mathbfss{C}}_l$ is $\mathbfss{C}_l$ (the estimator is unbiased), and the variance of a given map power spectrum is $2C_l^2/\nu$. The latter also follows immediately from squaring the definition Equation~\eqref{eq:Chatdef} and using Wick's theorem.

When $p=1$ this distribution is equivalent to a Gamma distribution. The Gamma distribution has non-zero skewness, kurtosis, and higher-order cumulants. For example, the dimensionless skewness is equal to $S = 2\sqrt{2/\nu}$, and the dimensionless excess kurtosis is $\kappa = 12/\nu$. It is easy to see that higher-order cumulants come with successive factors of $\nu^{-1/2}$; this is how the Central Limit Theorem manifests itself when $\nu \gg 1$, since these higher-order cumulants are all formally zero for a Gaussian field. From Equation~\eqref{eq:Chatdef} we see that $\nu$ counts the azimuthal modes contributing to a given wave number, such that the power spectrum estimator is the average of an increasingly large number of independent variables as $\nu$ increases.

The convergence of any probability distribution to a Gaussian can be established formally by showing that its characteristic function tends pointwise to that of a Gaussian (L\`{e}vy's continuity theorem, e.g.~Ref.~\cite{Durrett}). Inspection of Equation~\eqref{eq:wishchar} makes clear how this happens, where the convergence to a Gaussian can be established by Taylor-expanding around $\nu=\infty$. One can also establish Gaussianity from the density directly. In the $p=1$ case one first has to change variables to $\hat{C}_l = C_l(1 + n\sqrt{2/\nu})$ then take the limit $\nu \to \infty$ at fixed $n$, using Stirling's approximation to deal with the Gamma function in the denominator. Whichever way one chooses to go, the resulting limiting distribution is a product of Gaussians, each with mean $C_l$ and variance $2C_l^2/\nu$.

We close this section by noting that in general one has to be careful about assuming a Gaussian sampling distribution for power spectrum estimates, as this does not necessarily guarantee a Gaussian \emph{posterior} for parameters at the same level of accuracy. For example, if one has an informative and non-Gaussian prior on a parameter, the amount of constraining data required for the likelihood to dominate the posterior can be well above that needed to Gaussianize the likelihood. Secondly, and more importantly, posterior standard errors on parameters can shrink as fast as non-Gaussian contributions to the likelihood, meaning relative corrections from non-Gaussianity can remain important for posteriors even on small angular scales. In general these effects are most robustly studied with simulations. Ref.~\cite{2008PhRvD..77j3013H} provides a thorough discussion of residual non-Gaussian effects from failure of the Central Limit Theorem for Gaussian fields in the CMB context, finding that sufficiently broad multipole bins can reduce the error from approximating the likelihood as Gaussian to acceptable levels. We will return to this question in the cosmic shear context in Section~\ref{sec:quant}.

\subsection{Corrections to the tomographic shear power distribution from a non-Gaussian signal}
\label{subsec:Wc}

In the previous section we derived the distribution of angular power spectrum estimates assuming both signal and noise were Gaussian. In this section we will consider the more realistic scenario of a non-Gaussian signal, but will keep the assumption of Gaussian noise.

In the Gaussian case, the mode coupling integral in Equation~\eqref{eq:char_1} can be performed analytically due to the Gaussian form of the signal characteristic function $\phi_s$. The fully general non-Gaussian characteristic function of the signal has no analytic expression that can be written down, so to make progress we must restrict to \emph{weak} non-Gaussianity, in the sense that the higher-order cumulants of the signal are progressively suppressed. This is the Edgeworth expansion~\cite{1992ApJ...394L...5B, 1995ApJ...442...39J, 1995ApJ...443..479B, 1996MNRAS.283..983A, 1998A&AS..130..193B,2001MNRAS.328.1027T}, and truncating at leading order gives
\begin{align}
     \phi_s(\{\mathbf{k}_{lm}\}) = &\left(1 - i\frac{1}{6}\kappa^{ijk}_{\underline{lm}} k^i_{l_1 m_1} k^j_{l_2 m_2} k^k_{l_3 m_3} \vphantom{\frac12} \right.\nonumber \\
     &\left. \vphantom{\frac12} + \frac{1}{24} \kappa^{ijkm}_{\underline{lm}} k^i_{l_1 m_1} k^j_{l_2 m_2} k^k_{l_3 m_3}k^m_{l_4 m_4} - \frac{1}{72} \kappa^{ijk}_{\underline{lm}} \kappa^{lmn}_{\underline{l'm'}} k^i_{l_1 m_1} k^j_{l_2 m_2} k^k_{l_3 m_3}k^l_{l_1' m_1'} k^m_{l_2' m_2'} k^n_{l_3' m_3'} \right) \phi_s^G(\{\mathbf{k}_{lm}\}),
\label{eq:Edge}
\end{align}
where all repeated indices are summed over and $\phi_s^G$ is the Gaussian characteristic function. The compact notation $\underline{lm}$ denotes $l_1 m_1, l_2m_2, l_3m_3$ when attached to a three-point cumulant, and $l_1 m_1, l_2m_2, l_3m_3, l_4m_4$ when attached to a four-point cumulant. These are defined as
\begin{align}
    \kappa^{ijk}_{\underline{lm}} &\equiv \langle s^i_{l_1 m_1} s^j_{l_2 m_2} s^k_{l_3 m_3} \rangle \\
    \kappa^{ijkm}_{\underline{lm}} &\equiv \langle s^i_{l_1 m_1} s^j_{l_2 m_2} s^k_{l_3 m_3} s^m_{l_4 m_4} \rangle_c,
\end{align}
where the subscript $c$ denotes the fully connected part.

In Appendix~\ref{app:Edge} we discuss how the signal cumulants are related to those of the power spectrum estimates themselves. The covariance and leading order dimensionless skewness of $\hat{C}_l$ in the $p=1$ case are given by 
\begin{align}
\langle \Delta \hat{C}_l \Delta \hat{C}_{l'} \rangle &= \frac{2}{\nu}C_l^2 \delta_{l l'} + T_{ll'}, \label{eq:cov_mainbody} \\
\frac{\langle \Delta \hat{C}_l^3 \rangle}{\langle \Delta C_l^2 \rangle^{3/2}} &\approx \sqrt{\frac{8}{\nu}} + \sqrt{2\nu^3}\frac{\tilde{B}_{lll}}{C_l^3} + \frac{3\sqrt{2\nu}}{2}\frac{T_{ll}}{C_l^2}.
\end{align}
The skewness has a term coming from the Gaussian part of the signal ($\sqrt{8/\nu}$), plus a term proportional to the squared bispectrum of the signal $\tilde{B}_{lll}$, defined later in Equation~\eqref{eq:btilde_def}, and a term proportional to its trispectrum $T_{ll'}$. These latter two quantities are exactly those present in Equation~\eqref{eq:Edge}, and are also the leading order non-Gaussian corrections to the dimensionless kurtosis of $\hat{C}_l$. Indeed, as discussed in Appendix~\ref{app:Edge}, \emph{all} of the dimensionless cumulants of $\hat{C}_l$ have leading-order corrections from the trispectrum and squared bispectrum. Since these two quantities are of the same perturbative order this means that the cumulants of $\hat{C}_l$ do not progressively smaller as their order increases, and an Edgeworth expansion of the distribution of $\hat{C}_l$ around a Wishart will not work. This justifies our approach of expanding the underlying signal as an Edgeworth expansion, and then propagating to get the distribution of the power spectrum.

Substituting Equation~\eqref{eq:Edge} into Equation~\eqref{eq:char_1}, we immediately see that the leading order bispectrum term vanishes due to this term carrying an odd power of $k_{lm}$. The leading order contribution to the distribution is therefore from terms of order the trispectrum or bispectrum squared, whose computation formally requires going to third order in perturbation theory. We will therefore truncate the expansion in Equation~\eqref{eq:Edge} at the order shown. Since the probability density is linear in the characteristic function, we can simply add the trispectrum correction to that coming from bispectrum squared. We present the derivation of the characteristic function and the probability density of the power spectrum in Appendix~\ref{app:pne1}, with the final result given in Equations~\eqref{eq:finalpCl_tri_pne1} and~\eqref{eq:finalpCl_bisq_pne1}. We will simply quote the results here.

\subsubsection{Non-Gaussian power distribution from a signal with non-zero trispectrum}
\label{subsubsec:trispec}

In the $p=1$ case the distribution of $\hat{C}_l$ corrected by a signal trispectrum given in Equation~\eqref{eq:finalpCl_tri_pne1} reduces to
\begin{equation}
    p(\{\hat{C}_l\}) = \left\{\prod_{l=l_{\mathrm{min}}}^{l_{\mathrm{max}}}\Gamma[\hat{C}_l; \nu/2, \nu/(2C_l)]\right\} \left[1 + \frac{1}{8}\sum_{l_1, l_2} \nu_1 \nu_2 \frac{ T_{l_1 l_2}}{C_{l_1} C_{l_2}} \left( \frac{\Delta \hat{C}_{l_1}\Delta \hat{C}_{l_2}}{C_{l_1}C_{l_2}} - \frac{2}{\nu_1+2}\delta_{l_1 l_2} \frac{\hat{C}_{l_1}^2}{C_{l_1}^2}\right)\right],
    \label{eq:finalpCl_tri}
\end{equation}
where $T_{ll'}$ is the $m$-averaged version of the signal trispectrum that contributes to the power spectrum covariance, see Equation~\eqref{eq:cov_mainbody}. One can verify that the distribution Equation~\eqref{eq:finalpCl_tri} is correctly normalised, has a mean given by $\langle \hat{C}_l \rangle = C_l$ (unchanged from the Gamma case) and a covariance matrix given by $\langle \Delta \hat{C}_{l_1} \Delta \hat{C}_{l_2} \rangle = 2\delta_{l_1 l_2}C_{l_1}^2/\nu_1 + T_{l_1 l_2}$, as expected.

Despite possessing the correct normalisation, mean and covariance, Equation~\eqref{eq:finalpCl_tri} is not a valid probability density since it is not everywhere positive. This arises for sufficiently large $\hat{C}_l$. This is a ubiquitous feature of densities based on the Edgeworth expansion, and can be remedied by writing the correction as an exponential, i.e.~$1+x \approx e^x$. This is valid since the correction is assumed to perturbatively small.

The first correcting term in Equation~\eqref{eq:finalpCl_tri}, proportional to $\Delta \hat{C}_{l_1}\Delta \hat{C}_{l_2}$, can be written in a suggestive way. It is the leading-order term in the expansion of
\begin{equation}
    -\frac{1}{2} \sum_{l_1, l_2} \Delta \hat{C}_{l_1} \left( \frac{2C_{l_1}^2}{\nu_1}\delta_{l_1l_2} + T_{l_1 l_2} \right)^{-1} \Delta \hat{C}_{l_2},
    \label{eq:alachi2}
\end{equation}
where the notation $(A_{l_1 l_2})^{-1}$ refers to the $l_1,l_2$ element of the inverse of the matrix with elements $A_{l_1 l_2}$, as should be clear from context. Equation~\eqref{eq:alachi2} is a $\chi^2$-like functional having the inverse of the \emph{total} covariance matrix. Once the Gaussian part of this expansion is `available' from the zero-order Gamma term when the $\nu \to \infty$ limit is taken, we will be left a Gaussian possessing the correct inverse covariance matrix in its exponent. Similarly, the second term in Equation~\eqref{eq:finalpCl_tri} is the leading-order correction to the determinant pre-factor of a Gaussian due to the non-Gaussian covariance $T_{l_1 l_2}$. Thus, in the $\nu \to \infty$ limit, our distribution is the leading-order correction to a Gaussian due to $T_{l_1 l_2}$. This can also be seen directly from the characteristic function, Equation~\eqref{eq:char_with_tri_pne1}.

Considered as a function of the model power spectrum $C_{l}$ and the trispectrum $T_{l_1 l_2}$, Equation~\eqref{eq:finalpCl_tri} gives the likelihood function for fixed data $\hat{C}_l$. In the next section we will compute the correction to the likelihood numerically, but there are a few analytic results we can derive from this likelihood. Most significantly, we can compute its expected curvature around a fiducial power spectrum $C_{f,l}$ and non-Gaussian covariance $T_{f,l_1 l_2}$ i.e.~the Fisher matrix. Assuming a model with parameters $\theta_\alpha$, this is given by
\begin{align}
 F_{\alpha \beta} &= \frac{1}{2} \sum_{l=l_{\mathrm{min}}}^{l_{\mathrm{max}}} \nu \frac{1}{C_{f,l}^2}\frac{\partial C_l}{\partial \theta_\alpha}\frac{\partial C_l}{\partial \theta_\beta} - \frac{1}{4}\sum_{l_1,l_2=l_{\mathrm{min}}}^{l_{\mathrm{max}}} \nu_1\nu_2 \frac{T_{f,l_1 l_2}}{C_{f,l_1}^2C_{f,l_2}^2} \frac{\partial C_{l_1}}{\partial \theta_\alpha}\frac{\partial C_{l_2}}{\partial \theta_\beta} \nonumber \\
 &\approx \sum_{l_1,l_2=l_{\mathrm{min}}}^{l_{\mathrm{max}}} \frac{\partial C_{l_1}}{\partial \theta_\alpha} \left(\frac{2C_{f,l_1}^2}{\nu}\delta_{l_1l_2} + T_{f,l_1 l_2} \right)^{-1} \frac{\partial C_{l_2}}{\partial \theta_\beta},
 \label{eq:Fish_tri}
\end{align}
where the derivatives are evaluated at the fiducial model. Equation~\eqref{eq:Fish_tri} is the usual expression for the power spectrum Fisher matrix but with the proper non-Gaussian covariance matrix. This matches the expression for the Fisher matrix derived from a Gaussian likelihood with the covariance fixed to a fiducial model and including the non-Gaussian term, and tells us that to recover the full-sky Fisher matrix from a likelihood approximated by its Gaussian limit we have to \emph{fix} the Gaussian covariance to a fiducial model. This is the main result of Ref.~\cite{2013A&A...551A..88C}, and here we recover it in the presence of signal non-Gaussianity. It is interesting to note that in principle one can allow the non-Gaussian part of the covariance matrix, $T_{l_1 l_2}$, to depend on parameters in a Gaussian likelihood even when fixing the Gaussian part, as required to keep the approximation consistent. In our perturbative framework the additional information contributed by $T_{l_1 l_2}(\theta)$ is higher order, but whether this is consistent with the likelihood approximation for general $T_{l_1 l_2}$ is still an open question, and covered by neither our analysis nor that of Ref.~\cite{2013A&A...551A..88C}.

We note finally that we do not recover the field-level Fisher matrix from our power spectrum Fisher matrix, which now contains new terms due to non-Gaussianity. This is an expression of the fact that the power spectrum does not now contain all the information.

\subsubsection{Non-Gaussian power distribution from a signal with non-zero bispectrum}
\label{subsubsec:bispec}

We have argued that a consistent expansion of distribution of the power spectrum requires including terms of order the signal bispectrum squared. In the $p=1$ case, the correction given in Equation~\eqref{eq:finalpCl_bisq_pne1} reduces to
\begin{align}
    p(\{\hat{C}_l\}) = &\left\{\prod_{l=l_{\mathrm{min}}}^{l_{\mathrm{max}}}\Gamma[\hat{C}_l; \nu/2, \nu/(2C_l)]\right\} \nonumber \\
    & \times \left\{1 + \frac{1}{12}\sum_{l_1, l_2,l_3} \nu_1 \nu_2 \nu_3 \frac{\tilde{B}_{l_1l_2l_3}}{C^2_{l_1} C^2_{l_2} C^2_{l_3}}  \vphantom{\frac12} \left[\Delta \hat{C}_{l_1} \Delta \hat{C}_{l_2} \Delta \hat{C}_{l_3} - [3]\frac{2\delta_{l_1 l_2}}{(\nu_1+2)} \hat{C}_{l_1}^2\Delta \hat{C}_{l_3} + \frac{16\delta_{l_1 l_2}\delta_{l_2 l_3}}{(\nu_1+4)(\nu_1+2)}\hat{C}_{l_1}^3\right]\right\},
    \label{eq:finalpCl_bisq}
\end{align}
where $\tilde{B}$ is proportional to the square of the reduced bispectrum $b_{l_1l_2l_3}$ of the signal and is given by
\begin{equation}
    \tilde{B}_{l_1l_2l_3} = \frac{1}{4\pi} \ThreeJSymbol{l_1}{0}{l_2}{0}{l_3}{0}^2  b_{l_1l_2l_3}^2,
    \label{eq:btilde_def}
\end{equation}
with the quantity in parentheses a Wigner $3j$ symbol. The notation $[n]$ in Equation~\eqref{eq:finalpCl_bisq} refers to the $n$ terms that follow from symmetry considerations. It is straightforward to verify that the correction leaves the normalisation, mean, and covariance matrix unchanged, but corrects the three-point function by a term equal to $4\tilde{B}_{l,l',l''}$, as expected from Equation~\eqref{eq:skewofCl}.

One can verify that the Fisher matrix is unaffected by the squared bispectrum correction at leading order. In other words, the effect of $\tilde{B}_{l_1 l_2 l_3}$ is to modify the skewness of the likelihood, while leaving its mean and variance approximately unaffected. 

As claimed in Section~\ref{app:weakNG}, the correction to the density is $\mathcal{O}(\tilde{B}/C_l^3)$, the same order as the trispectrum correction. This justifies our decision to include terms of order the bispectrum squared.

In summary, the leading-order correction to the distribution of power spectrum estimates from signal non-Gaussianity is
\begin{align}
    p(\{\hat{C}_l\} \vert \{C_l,T_{l,l'}, \tilde{B}_{l,l',l''}\}) &= \left\{\prod_{l=l_{\mathrm{min}}}^{l_{\mathrm{max}}}\Gamma[\hat{C}_l; \nu/2, \nu/(2C_l)]\right\} \left\{1 + \frac{1}{8}\sum_{l_1, l_2} \nu_1 \nu_2 \frac{ T_{l_1 l_2}}{C_{l_1} C_{l_2}} \left( \frac{\Delta \hat{C}_{l_1}\Delta \hat{C}_{l_2}}{C_{l_1}C_{l_2}} - \frac{2}{\nu_1+2}\delta_{l_1 l_2} \frac{\hat{C}_{l_1}^2}{C_{l_1}^2}\right) \right. \nonumber \\
    &\left. + \frac{1}{12}\sum_{l_1, l_2,l_3} \nu_1 \nu_2 \nu_3 \frac{\tilde{B}_{l_1l_2l_3}}{C^2_{l_1} C^2_{l_2} C^2_{l_3}}  \vphantom{\frac12} \left[\Delta \hat{C}_{l_1} \Delta \hat{C}_{l_2} \Delta \hat{C}_{l_3} - [3]\frac{2\delta_{l_1 l_2}}{(\nu_1+2)} \hat{C}_{l_1}^2\Delta \hat{C}_{l_3} + \frac{16\delta_{l_1 l_2}\delta_{l_2 l_3}}{(\nu_1+4)(\nu_1+2)}\hat{C}_{l_1}^3\right]\right\}.
    \label{eq:fullcorrection}
\end{align}

Equation~\eqref{eq:fullcorrection} is the first major result of this work. It is the leading-order likelihood function for the power spectrum that incorporates a non-Gaussian signal. Unlike the commonly assumed Gaussian approximation, this distribution applies equally well to the largest angular scales. This distribution has the correct normalization, mean, covariance, and three-point function. In Appendix~\ref{app:pne1} we provide the analogous expression in the case of multiple redshift bins, in which case the distribution is a perturbation around a Wishart distribution for positive definite matrices. A practical application of the distribution may require ensuring positivity by taking the logarithm, but in practice we have not found this to be necessary. In Section~\ref{sec:quant} we present numerical results for the transition of the power spectrum distribution from Wishart to post-Wishart (our Equation~\eqref{eq:fullcorrection}) and finally to Gaussian as the number of degrees of freedom increases.

We close this section by noting that while we have restricted ourselves to spin-0 fields (such as lensing convergence or projected galaxy number counts) for simplicity, our formalism may be easily generalized to the spin-2 case (i.e.~weak lensing shear), as mentioned at the beginning of Section~\ref{sec:likes}. Since we work in harmonic space on the full sky, one can simply work with the $E$ and $B$ multipoles derived in the usual way from the shear field. This augments the data vector to have length $2\times N_z$, but all the matrix expressions presented here can be used after making this modification. Redshift bin indices must now be interpreted as also running over $E$ and $B$ indices, and the appropriate cross-bispectra and trispectra must be used (many of which will be zero due to parity invariance). 

On the cut sky our expression must be modified to account for the mask, which we have not attempted to include. This induces additional mode mixing (as well as the usual $E$/$B$ ambiguity), which could possibly be included using the techniques of either Ref.~\cite{2008PhRvD..77j3013H} or Ref.~\cite{2020MNRAS.491.3165U}. We defer this to a future work.

\subsection{The transition to Gaussianity}
\label{subsec:CLT}

A key goal of this paper is to understand and quantify how efficiently the Central Limit Theorem (CLT) Gaussianizes the distribution of $\hat{C}_l$ in the presence of signal non-Gaussianity. One could well imagine a situation for example where the strength of non-Gaussianity is increasing with $l_{\mathrm{max}}$ faster than the CLT is able to Gaussianize, in which the Gaussian limit may \emph{never} be achieved. In this section we use our new distribution to study how the CLT operates in detail. We specialise to the single redshift bin case, $p=1$.

Convergence of a probability density to a limiting distribution can be established by studying the pointwise convergence of its characteristic function. We have already computed this, and we will additionally change variables from $\hat{C}_l$ to the `standardised' variable $\hat{Z}_{l} \equiv \Delta \hat{C}_l/\sqrt{2C_l^2/\nu}$. The characteristic function of the set of $\hat{Z}_l$ variables is
\begin{align}
\phi_{\{\hat{Z}_l\}}(\{J_l\}) &= \left[\prod_{l=l_{\mathrm{min}}}^{l_{\mathrm{max}}} e^{iJ_l\sqrt{\nu/2}} (1 + iJ_l \sqrt{2/\nu})^{-\nu/2}\right] 
    \left[1 - \frac{1}{2}\sum_{l_1,l_2}\varepsilon_{l_1 l_2} \frac{J_{l_1}J_{l_2}}{(1 + iJ_{l_1}\sqrt{2/\nu_1})(1 + iJ_{l_2}\sqrt{2/\nu_2})} \right. \nonumber \\
    & \left. + \frac{i}{6}\sum_{l_1,l_2,l_3}\tilde{\varepsilon}_{l_1 l_2 l_3} \frac{J_{l_1}J_{l_2}J_{l_3}}{(1 + iJ_{l_1}\sqrt{2/\nu_1})(1 + iJ_{l_2}\sqrt{2/\nu_2})(1 + iJ_{l_3}\sqrt{2/\nu_3})}\right],
    \label{eq:charz}
\end{align}
where we have defined the quantities
\begin{align}
\varepsilon_{l_1 l_2} &\equiv \frac{T_{l_1 l_2}}{\sqrt{2C_{l_1}^2/\nu_1}\sqrt{2C_{l_2}^2/\nu_2}}, \label{eq:epsdef}\\
\tilde{\varepsilon}_{l_1 l_2 l_3} &\equiv \frac{4\tilde{B}_{l_1 l_2 l_3}}{\sqrt{2C_{l_1}^2/\nu_1}\sqrt{2C_{l_2}^2/\nu_2}\sqrt{2C_{l_3}^2/\nu_3}}.
\end{align}
The quantity $\varepsilon_{l_1 l_2}$ has the simple interpretation that it is the relative correction to the power spectrum covariance matrix from non-Gaussianity. Its diagonal elements can also be interpreted as the correction to the `effective degrees of freedom' through $\nu_{\mathrm{eff}} = \nu/(1+\varepsilon_{ll})$ such that the variance of the power spectrum is $2C_l^2/\nu_{\mathrm{eff}}$, although simply making this correction in the likelihood is not sufficient as it does not capture, among other things, mode coupling.

Consider now taking every $\nu \gg 1$, i.e.~consider the limit $l_{\mathrm{min}} \gg 1$. The zero-order prefactor in Equation~\eqref{eq:charz} converges to a Gaussian, with the `linear' part of the exponent getting cancelled by the complex exponential already present. Likewise, all the denominators in the non-Gaussian correction terms tend to unity at the same rate, leaving a quadratic correction from the trispectrum and a cubic correction from the bispectrum. In this limit, the correction is of the Edgeworth form. At the order to which we work, we may just as well replace the quadratic term by a Gaussian using $e^{-\lambda^2} \approx 1 - \lambda^2$, in which case the trispectrum term is absorbed into the covariance matrix of the Gaussian prefactor. This leaves the correction in the $l_{\mathrm{min}} \gg 1$ limit as
\begin{equation}
\phi_{\{\hat{Z}_l\}}(\{J_l\}) \approx \exp\left[-\frac{1}{2}\sum_{l_1 l_2} J_{l_1} (\delta_{l_1 l_2} + \varepsilon_{l_1 l_2}) J_{l_2}\right] \left(1 + \frac{i}{6}\sum_{l_1,l_2,l_3}\tilde{\varepsilon}_{l_1 l_2 l_3} J_{l_1}J_{l_2}J_{l_3}\right).
\end{equation}
For this to be valid, we clearly need $\sum_{l_1 l_2} J_{l_1} J_{l_2} \varepsilon_{l_1 l_2} \ll 1$. A necessary condition for this is that every element of $\varepsilon$ is `small' in a sense that depends on the quantity of interest. For example, if one was only interested in the marginal distribution of a single $\hat{C}_l$, it would be sufficient to demand that only a single $\varepsilon_{ll} \ll 1$.

The dominant effect of the trispectrum in this limit is clearly a change to the covariance matrix. Residual `post-Gaussian' effects come from the squared bispectrum through the quantity $\tilde{\varepsilon}_{l_1 l_2 l_3}$ affecting the three-point function of the power spectrum. These are negligible when $\tilde{\varepsilon}_{l_1 l_2 l_3} \ll 1$ at all multipoles.

One guaranteed way to make both $\varepsilon$ and $\tilde{\varepsilon}$ small is to increase $C_l$ at fixed $T_{l_1 l_2}$ or fixed $\tilde{B}_{l_1 l_2 l_3}$. This may be achieved by simply increasing the (assumed Gaussian) noise power, since $C_l$ here is the \emph{total} power spectrum. This is equivalent to saying that the noise `Gaussianizes' the distribution. One must be careful however to check this Gaussianization also operates on not just the likelihood but the posterior, as we will do shortly.

The quantity $\varepsilon_{l_1 l_2}$ is known to be significant on non-linear scales and is routinely measured in simulations (usually in a binned form, see Figure~\ref{fig:covdiff} for a related quantity). In contrast, the size of $\tilde{\varepsilon}_{l_1 l_2 l_3}$ is less obvious. We can make an order-of-magnitude estimate of its size using the asymptotic limit of the Wigner-$3j$ symbol~\cite{ponzano, 2011JCAP...02..015B} 
\begin{equation}
    \lim_{\nu_1, \nu_2, \nu_3 \to \infty} \ThreeJSymbol{l_1}{0}{l_2}{0}{l_3}{0}^2 = \frac{1}{2\pi A(L_1,L_2,L_3)},
\end{equation}
where $L \equiv l+1/2$ and $A(L_1,L_2,L_3)$ is the area of the triangle having side lengths $L_1$, $L_2$, $L_3$. The squared $3j$ symbol thus decays as $1/\nu^2$ for large $\nu = \mathrm{min}(\nu_1, \nu_2, \nu_3)$. A necessary condition for negligible non-Gaussianity is therefore
\begin{equation}
    \left[\frac{\sqrt{\nu_1 \nu_2 \nu_3/2}}{4\pi^2A(L_1,L_2,L_3)}\right]^{1/2} \frac{\lvert b_{l_1l_2l_3}\rvert}{\sqrt{C_{l_1}C_{l_2}C_{l_3}}} \ll 1
    \label{eq:bsqlim}
\end{equation}
when $l_{\mathrm{min}} \gg 1$. The pre-factor on the left-hand side of Equation~\eqref{eq:bsqlim} is a weak function of $\nu$, and is typically $\sim 10^{-2}$ for the relevant scales and for most triangle configurations. The reduced bispectrum for weak lensing has been measured for a few triangle configurations from $N$-body simulations in Ref.~\cite{2020MNRAS.493.3985M}, who find that it \emph{decays} with $l$ for most configurations, with the slowest decay being a roughly $l^{-1}$ drop-off in their `squeezed' configuration. This suggests that for the lensing convergence, the bispectrum-squared term should become rapidly negligible with $l$. At $l \gtrsim 2000$ where shape noise starts to dominate the power spectrum, \cite{2020MNRAS.493.3985M} find that the reduced bispectrum for an equilateral configuration is $\sim 10^{-17}$, and therefore the left-hand side of Equation~\eqref{eq:bsqlim} is $\sim 10^{-5}$. At $l=100$ we have $C_l \sim 10^{-9}$ and $b_{lll} \sim 10^{-17}$, so the left-hand side of Equation~\eqref{eq:bsqlim} is again $\sim 10^{-5}$. For a squeezed shape with $l_1 = 50$, $l_2=l_3=2000$, \cite{2020MNRAS.493.3985M} finds $b \sim 10^{-16}$. We find that $C_{l_1} ~\sim 10^{-8}$, and the left-hand side of Equation~\eqref{eq:bsqlim} is $\sim 10^{-4} $ i.e.~slightly more significant. A typical element of $\tilde{\varepsilon}_{l_1 l_2 l_3}$ is therefore around $10^{-9}$.

In practice the quantity that should be small is the \emph{sum} of $\varepsilon$ across the scales most relevant to the quantity of interest. The order-of-magnitude analysis above suggests that for a small subset of the $C_l$ the correction from non-Gaussianity is negligible, but if the number of relevant scales is large (as might be the case for massive data compression in order to infer cosmological parameters) then the conditions on $\varepsilon$ are more strict. Assuming an approximately constant $\tilde{\varepsilon}_{l_1 l_2 l_3}$, using all available scales up to $l_{\mathrm{max}}$ would require $\tilde{\varepsilon}_{l_1 l_2 l_3} \ll l_{\mathrm{max}}^{-3}$ for example. For $\tilde{\varepsilon}_{l_1 l_2 l_3} \sim 10^{-9}$ corrections become large around $l_{\mathrm{max}} \approx 1000$. This is within the range of scale useable angular scales for Stage-IV lensing surveys, which motivates the more quantitative study with simulations in Section~\ref{sec:quant}. 
%

\subsubsection{Non-Gaussian effects on posterior distributions}
\label{subsubsec:post}

The sampling distribution of $\hat{C}_l$ itself is usually not of interest. What matters more is the \emph{posterior} distribution on the $C_l$, or on model parameters that generate the $C_l$. In this section we study analytically how non-Gaussianity affects the posterior, using our corrected likelihood Equation~\eqref{eq:fullcorrection}.

First we study the joint posterior distribution of the set of theory $C_l$. This is simply given by $p(\{C_l\} \vert \{\hat{C}_l\}) \propto p(\{\hat{C}_l\} \vert \{C_l\})\pi(\{C_l\})$, where $\pi(\{C_l\})$ is the prior. Natural priors for the $C_l$ are the uniform prior, which isolates the dependence on the likelihood, and the Jeffreys prior, which is uninformative. For a Gaussian likelihood the two coincide, but this is not the case for either the Gamma likelihood nor our Equation~\eqref{eq:fullcorrection}. 

The Jeffreys prior is given by the square-root of the determinant of the Fisher matrix. We have computed this for our likelihood, and at leading order we have
\begin{equation}
\pi(\{C_l\}) \propto \left(\prod_{l=l_{\mathrm{min}}}^{l_{\mathrm{max}}} C_l^{-1} \right)\left(1 - \frac{1}{4}\sum_l \nu \frac{T_{ll}}{C_l^2} \right)
\end{equation}
The zero-order posterior for each $C_l$ is thus an Inverse-Gamma distribution, denoted by $\Gamma^{-1}$ given by
\begin{equation}
p_{\Gamma^{-1}}(C_l \vert \hat{C}_l) = \frac{(\nu \hat{C}_l/2)^{\nu/2}}{\Gamma(\nu/2)} \, C_l^{-\nu/2 -1} \exp\left(-\frac{\nu\hat{C}_l}{2C_l}\right).
\end{equation}
Weak signal non-Gaussianity perturbs away from this zero-order distribution.

In what follows we will only include trispectrum terms, which we have found to be the dominant source of non-Gaussianity in the power spectrum distribution. Substituting in the Jeffreys prior and normalizing, the correction to the Inverse-Gamma posterior is
\begin{align}
p(\{C_l\} \vert \{\hat{C}_l\}) = &\left[\prod_{l=l_{\mathrm{min}}}^{l_{\mathrm{max}}} \Gamma^{-1}(C_l; \nu/2, \nu C_l/2)\right] \nonumber \\
& \times \left\{1 + \frac{1}{4}\sum_{l_1 l_2}\hat{\varepsilon}_{l_1 l_2}\left[ \frac{\hat{C}_{l_1}\hat{C}_{l_2}}{C_{l_1}C_{l_2}} \left(\frac{\Delta \hat{C}_{l_1} \Delta\hat{C}_{l_2}}{C_{l_1}C_{l_2}} - \frac{2\delta_{l_1 l_2}}{\nu_1+2}\frac{\hat{C}_{l_1}^2}{C_{l_1}^2} - \frac{2\tilde{\alpha}\delta_{l_1 l_2}}{\nu_1}\right) + \frac{\delta_{l_1l_2}[2\tilde{\alpha}(\nu_1+2) - 4] - 4}{\nu_1 \nu_2} \right]\right\}
\end{align}
where $\hat{\varepsilon}$ is defined as in Equation~\eqref{eq:epsdef} but with $\hat{C}_l$ in place of $C_l$, and $\tilde{\alpha}$ is a book-keeping parameter that keeps track of the influence of the correction to the Jeffreys prior; $\tilde{\alpha} = 1$ represents the first-order-corrected Jeffreys prior, and $\tilde{\alpha} = 0$ uses only the zero-order Jeffreys prior.

Corrections to the posterior moments from $T_{l_1l_2}$ can now be computed easily using the properties of the Inverse-Gamma distribution. For example, the mean of a single $C_L$ after marginalising over all other $C_l$ receives a fractional correction of order $\mathcal{O}(\hat{\varepsilon}/\nu)$, i.e.~ suppressed at large $\nu$. The leading-order fractional correction to the marginal variance of a single $C_L$ is $\hat{\varepsilon}_{LL}$, as expected, and is entirely captured by a Gaussian having the full non-Gaussian covariance matrix. The leading-order `post-Gaussian' correction from signal non-Gaussianity is $\mathcal{O}(\hat{\varepsilon}/\nu)$. Note the variance under a Gaussian approximation also differs from that of the Inverse-Gamma by a fractional amount $\mathcal{O}(1/\nu)$ when $\nu$ is finite. Our formalism is valid at all orders in $1/\nu$, but does assume small $\varepsilon$. Corrections to the mean and variance not captured by a Gaussian with non-Gaussian covariance matrix are thus $\mathcal{O}(\hat{\varepsilon}/\nu)$. It turns out that these are also the corrections to the \emph{conditional} mean and variance when all but one $C_l$ are fixed to their measured values, with the mean correction further suppressed if $\tilde{\alpha} = 1$.

To study the Gaussian limit in the $C_l$ posterior, we write $C_l = \hat{C}_l(1 + n\sqrt{2/\nu})$ and take the $\nu \to \infty$ limit. The leading-order term, quadratic in $n$, is precisely that specified by a Gaussian with non-Gaussian covariance. Leading-order post-Gaussian terms at fixed $n$ are $\mathcal{O}(\varepsilon/\sqrt{\nu})$, i.e.~larger than the correction to the mean and variance. This correction is a cubic polynomial in $n$, suggesting that it corresponds to a change in the skewness of the posterior $C_l$, not captured by a Gaussian. It turns out that this correction can be accounted for by a slight modification to the Gaussian where we replace $T_{l_1 l_2}$ in the covariance matrix with $T_{l_1 l_2}(\hat{C}_{l_1}^2/C_{l_1}^2)(\hat{C}_{l_2}^2/C_{l_2}^2)$, but in practice this post-Gaussian term is small and the correction does not lead to noticeable improvement. Note that we have neglected the bispectrum-squared term in this derivation, which will also modify the $C_l$ skewness.

%
%

We can also study the distribution of an amplitude parameter by setting $C_l = A C_{0,l}$ for some fiducial power spectrum $C_{0,l}$ (we neglect noise here for simplicity). We find a fractional correction to the mean of order $\mathcal{O}(\langle \varepsilon \rangle /\sum \nu)$, where we define the $l$-averaged quantity
\begin{align}
\langle \varepsilon \rangle &\equiv \frac{\sum_{l_1 l_2} \sqrt{\nu_1 \nu_2} \varepsilon_{l_1 l_2}}{\sum_l \nu} \nonumber \\
&= \frac{\sum_{l_1 l_2} \nu_1 \nu_2 \frac{T_{l_1 l_2}}{2C_{l_1}C_{l_2}}}{\sum_l \nu},
\label{eq:varepsilon_av}
\end{align}
with $\varepsilon$ evaluated at the fiducial model. $\langle  \varepsilon \rangle$ has the simple interpretation that it is the leading-order fractional correction to the posterior parameter variance from the non-Gaussian covariance term. Similarly, the leading-order post-Gaussian fractional correction to the variance of $A$ is $\mathcal{O}(\langle \varepsilon \rangle/\sum \nu)$. Corrections to the dimensionless skew of $A$ are $\mathcal{O}(\langle \varepsilon \rangle/\sqrt{\sum \nu})$, as are corrections to the pointwise posterior distribution of $A$.
%
%
%

To summarise this section, the biggest effect of a signal trispectrum is to correct the posterior variance of parameters through the non-Gaussian covariance term. This is typically of order $\langle \varepsilon \rangle$, where the average runs over the scales of interest. We require this correction to be small in order for our perturbative series to converge, i.e. the validity of our distributions require $\langle \varepsilon \rangle \ll 1$. A good rule of thumb is that we require the correction to the Fisher information from signal non-Gaussianity to be much less than unity. The broadening of parameter errors is mostly accounted for by a Gaussian likelihood with non-Gaussian covariance - the genuinely new terms we have computed here are `post-Gaussian', and give a skewness and fractional pointwise correction to the probability density which are both of order $\langle \varepsilon \rangle /\sqrt{\sum \nu}$ at large $\nu$, which by Equation~\eqref{eq:dimlessskewCl} is also the leading correction to the dimensionless skewness of the measured power spectrum. At large $\nu$ this is strongly suppressed due to the Central Limit Theorem if $\langle \varepsilon \rangle $ does not grow faster than $\sqrt{\sum \nu}$. In Section~\ref{sec:quant} we quantify this with simulations. Corrections due to the squared bispectrum also affect this skewness. Corrections to the posterior mean and variance are smaller, of order $\langle \varepsilon \rangle /\sum \nu$ at large $\nu$, which is of order the correction to the dimensionless kurtosis of the measured power spectrum.

\subsection{Bandpowers}
\label{subsec:binning}

It is usually advantageous to bin the power spectrum in angular scale in order to reduce dimensionality of the required covariance matrix, account for mode mixing due to the survey mask, or to speed up likelihood calculations. Binning has a strong effect on the apparent impact of non-Gaussianity in the covariance matrix, since the Gaussian term will be suppressed due to the increasing mode count while the non-Gaussian term will stay roughly constant.

Our binning scheme is defined by
\begin{equation}
\hat{C}_b = \frac{\sum_{l \in b} W_l \hat{C}_l}{\sum_{l \in b} W_l},
\end{equation}
for a window function $W_l$, where $b$ labels the bin. For narrow bins, an optimal binning scheme would weight with the inverse covariance of $\hat{C}_l$. To avoid dependence on the model we weight by the mode count and take $W_l = 2l + 1$. We explore different binning choices in Appendix~\ref{app:dlog0p2}. 

Assuming $C_l$ is roughly constant over the extent of the bin, the zero-order characteristic function of the $\hat{C}_b$ is
\begin{align}
    \phi_{\{ \nu_b \hat{C}_b\}}(\{ J_b \}) &= \prod_{b} \prod_{l \in b}  (1 + 2iJ_b C_l)^{-\nu/2} \nonumber \\
    &\approx \prod_b (1 + 2iJ_b C_b)^{-\nu_b/2},
\end{align}
where $\nu_b \equiv \sum_{l \in b} \nu$. In other words, the zero-order distribution of the binned power spectra is approximately a Gamma distribution. In practice we can eliminate the need to approximate a constant $C_l$ over the bin by working with $\hat{C}_l/C_l$, but for sufficiently narrow bins the above approximation is accurate.

In the presence of signal non-Gaussianity, we can make a similar approximation that $C_l$ is flat across each bin and simply replace $l_{1,2,3}$ with $b_{1,2,3}$ and $\nu_{1,2,3}$ with $\nu_{b_1,b_2,b_3}$ everywhere in the likelihood. Quantities such as $T_{l_1 l_2}$ and $\tilde{B}_{l_1 l_2 l_3}$ can be replaced with their binned versions.

Since the number of modes in a bin scales with its width $\Delta l_b$, the fractional correction to the covariance scales as $\varepsilon_{b b} \propto \Delta l_{b}$. The number of terms in the sum over $\varepsilon_{b_1 b_2}$ in the denominator of $\langle \varepsilon \rangle$ scales between $1/\Delta l_b^2$ for non-sparse $\varepsilon_{b_1 b_2}$ and $1/\Delta l_b$ for diagonal $\varepsilon_{b_1 b_2}$. Therefore, corrections from the signal trispectrum scale somewhere between $\langle \varepsilon \rangle \propto \Delta l_b$ for diagonal $\varepsilon$ and $\langle \varepsilon \rangle \approx \mathrm{constant}$ for non-sparse $\varepsilon$. In practice the $\varepsilon$ is not diagonal, and so we can expect our likelihood corrections to be only mildly sensitive to the choice of binning, with the generic feature that the corrections are larger for broader bins. We confirm this in Appendix~\ref{app:dlog0p2}.

\section{Quantifying corrections from non-Gaussianity}
\label{sec:quant}

In the previous section we derived the leading order correction to the probability density of power spectrum measurements in the presence of signal non-Gaussianity. Considered as a function of the unknown true power spectrum $C_l$, the signal trispectrum $T_{ll'}$, and the squared bispectrum $\tilde{B}_{ll'l''}$, this distribution gives the likelihood function. It is the impact of non-Gaussianity on the likelihood (or posterior) that is of most interest to us here.

We have seen that the Fisher information contributed by the parameter dependence of $T_{ll'}$ and $\tilde{B}_{ll'l''}$ is formally of higher order in the (assumed weak) non-Gaussianity, so in this section we will fix these to a fiducial model. Our approach is to use approximate simulations to compute these quantities in a fiducial cosmology, but we could just as well compute them using the halo model or fitting functions. Once we have computed these non-Gaussian signal cumulants, we can use Equation~\eqref{eq:fullcorrection} to compute the correction to the likelihood of the $C_l$, or any parameters on which $C_l$ depends.

\subsection{Simulations}
\label{subsec:simtest}

To compute $T_{ll'}$ and $\tilde{B}_{ll'l''}$ we create a large suite of lognormal lensing convergence maps using the \textsc{flask} software package~\cite{2016MNRAS.459.3693X}. Lognormal fields are known to give a reasonable approximation for the covariance of lensing two-point functions~\cite{2002ApJ...571..638T, 2011A&A...536A..85H}, although their ability to predict configurations of the bispectrum is less well explored~\cite{2012A&A...540A...9M}. We will cross-validate our results with a small sample of lensing maps from ray-traced from $N$-body simulations.

We generated $10^5$ lognormal convergence maps with a fixed power spectrum generated with \textsc{camb}~\cite{2000ApJ...538..473L, 2012JCAP...04..027H}. Our fiducial galaxy sample is modelled with a Euclid-like lensing sample in mind. We assume a single broad redshift distribution for lensing sources, peaked around $z\approx 0.9$, having the same functional form as Ref.~\cite{2020A&A...642A.191E}, with cosmological parameters set to the best-fit values of Ref.~\cite{2020A&A...641A...6P}. The alternative choice of a narrow source redshift bin peaked at $z_s = 0.325$, corresponding to the lowest redshift bin of a Euclid-like survey with 10 tomographic bins, is explored in Appendix~\ref{app:zs6}. We compute the power spectrum to $l_{\mathrm{max}} = 6143$, applying non-linear and baryon feedback corrections according to the default specifications of Ref.~\cite{2021MNRAS.502.1401M}. For simplicity we do not include intrinsic alignments in the model; this will not affect our results. The `shift parameter' corresponding to the (negative) minimum value of the convergence fed to \textsc{flask} is set to $0.01214$, which was chosen by hand to get a good match to $N$-body simulations, and is comparable to the value found in Ref.~\cite{2021MNRAS.508.3125F} by fitting the 1-point kurtosis of the convergence. Convergence fields are rendered on a \textsc{healpix}\footnote{\url{https://healpix.sourceforge.io}}~\cite{2005ApJ...622..759G} grid with $N_{\mathrm{side}} = 2048$. In our actual likelihood computations we only use scales down to $l_{\mathrm{max}} = 1000$, which is sufficiently small to ensure the input power spectrum is recovered with negligible bias\footnote{The exponentiation of the Gaussian field needed to produce a lognormal field transfers power to sub-grid scales where it cannot be recovered from the discrete \textsc{healpix} grid. This causes a negative bias in the power spectrum on all scales that increases in severity as the Nyquist frequency $l_{\mathrm{Ny}} \approx 3N_{\mathrm{side}} - 1$ is approached.}, but sufficiently large to cover a realistic range of scales for lensing surveys. We apply the \textsc{healpix} pixel window function to the map to mimic the effect of coarse binning of a higher resolution galaxy catalogue.
%

To check that the lognormal fields are producing sensible results for the signal cumulants, we compare the convergence statistics with those from a suite of ray-traced $N$-body simulations from Ref.~\cite{2017ApJ...850...24T}. These simulations assume a thin source redshift plane at $z_s = 1.0334$, but the resulting lensing power spectrum only differs from a fiducial $C_l$ by a few percent across most scales. 

\begin{figure}
\centering
\includegraphics[width=0.8\columnwidth]{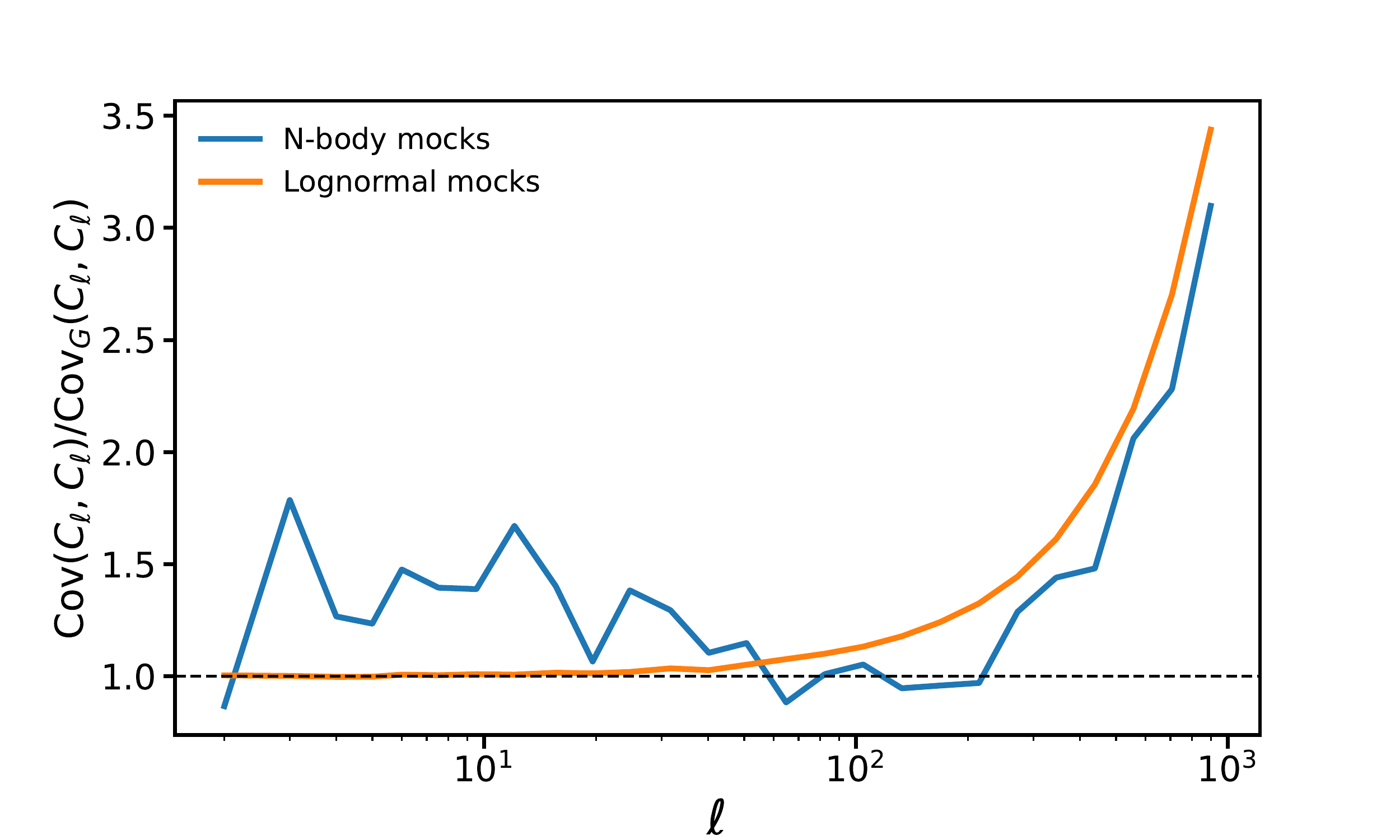}
\caption{Ratio of the diagonal elements of the $C_l$ covariance matrix to that of Gaussian fields for our lognormal mocks (orange) and the ray-traced $N$-body simulations of Ref.~\cite{2017ApJ...850...24T} (blue). We have binned the power spectra with $\Delta \log_{10} l = 0.1$. Ref.~\cite{2017ApJ...850...24T} speculate that the excess variance at large angular scales in the $N$-body mocks is due to the finite thickness of their lensing shells.}
\label{fig:covdiff}
\end{figure}

\begin{figure}
\centering
\includegraphics[width=0.8\columnwidth]{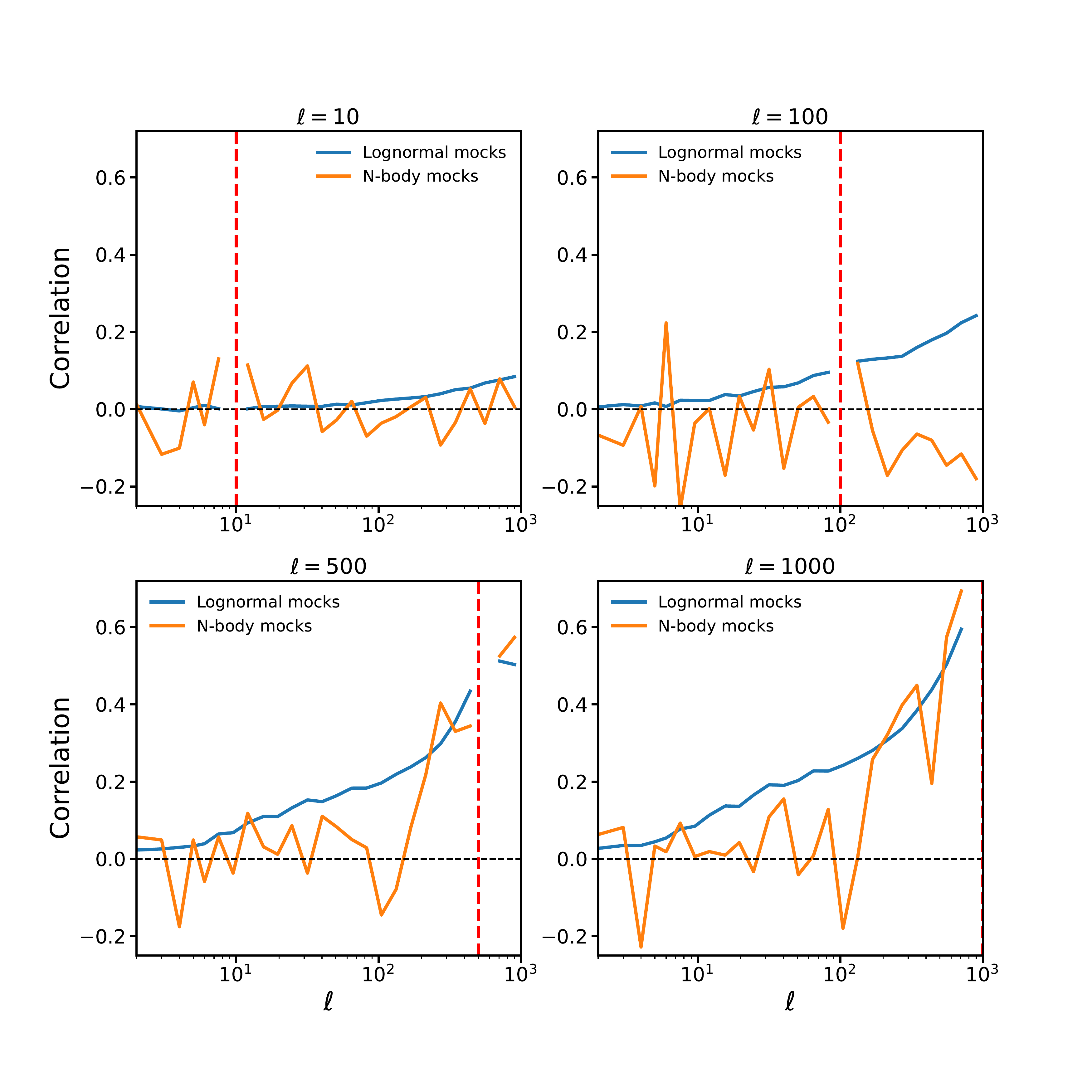}
\caption{Slices through the correlation matrix for a few multipole bins containing the $l$ values indicated above each panel. The diagonal elements (unity) has been omitted for clarity, and are indicated by the red dashed vertical lines.}
\label{fig:corrmat_slice}
\end{figure}

In Figures~\ref{fig:covdiff} and \ref{fig:corrmat_slice} we compare the covariance matrix of the measured $\hat{C_l}$ from the lognormal mocks with those of the $N$-body mocks. Since the latter contains much fewer realizations ($\sim 100$) the recovered statistics are much noisier, but a meaningful comparison is still possible. Figure~\ref{fig:covdiff} shows the ratio of the diagonal elements of the covariance matrix compared with a Gaussian prediction based on the input power spectrum (we have binned the $\hat{C}_l$ to emphasise the differences). The agreement between the two is reasonable, with the $N$-body mocks having excess variance on large scales, as discussed in Ref.~\cite{2017ApJ...850...24T}. The lognormal mocks overestimate the variance on small scales, but since we are only seeking approximations to the trispectrum here the difference is sufficiently small. The non-Gaussian contribution to the covariance dramatically kicks up when $l > 200$, due to the signal trispectrum. Note that our lognormal maps are on the full sky and are produced by a local non-linear transform of the linear field, so do not capture the super-sample covariance, but the agreement with the $N$-body mocks is nonetheless acceptable.


In Figure~\ref{fig:corrmat_slice} we plot slices through the correlation matrix. The agreement of the lognormal mocks with the $N$-body mocks is not perfect, but acceptable. Above $l\approx 200$ mode coupling in the $N$-body mocks kicks up dramatically, much in the same way as in the diagonal elements. The lognormal mocks do not capture this well, instead predicting a smooth increase, but the agreement is within an order of magnitude and acceptable.

To measure the bispectrum from the simulations we use the estimator
\begin{equation}
\hat{b}_{l_1 l_2 l_3} = \frac{1}{N_{l_1 l_2 l_3}} \sum_{m_1=-l_1}^{l_1} \sum_{m_2=-l_2}^{l_2} \sum_{m_3=-l_3}^{l_3} \ThreeJSymbol{l_1}{m_1}{l_2}{m_2}{l_3}{m_3} a_{l_1 m_1} a_{l_2 m_2} a_{l_3 m_3}, \label{eq:bispecest}
\end{equation}
where
\begin{equation}
N_{l_1 l_2 l_3} = \sqrt{\frac{\nu_1 \nu_2 \nu_3}{4\pi}} \ThreeJSymbol{l_1}{0}{l_2}{0}{l_3}{0},
\end{equation}
which is optimal for Gaussian fields. We use the Python package {\tt spherical}\footnote{\url{https://pypi.org/project/spherical/}} to precompute the Wigner $3j$ symbols to allow for rapid summation over the azimuthal wave numbers in Equation~\eqref{eq:bispecest}, but the implementation is slow for $l>100$ so we limit our bispectrum measurements to this maximum multipole.

\begin{figure}
\centering
\includegraphics[width=0.8\columnwidth]{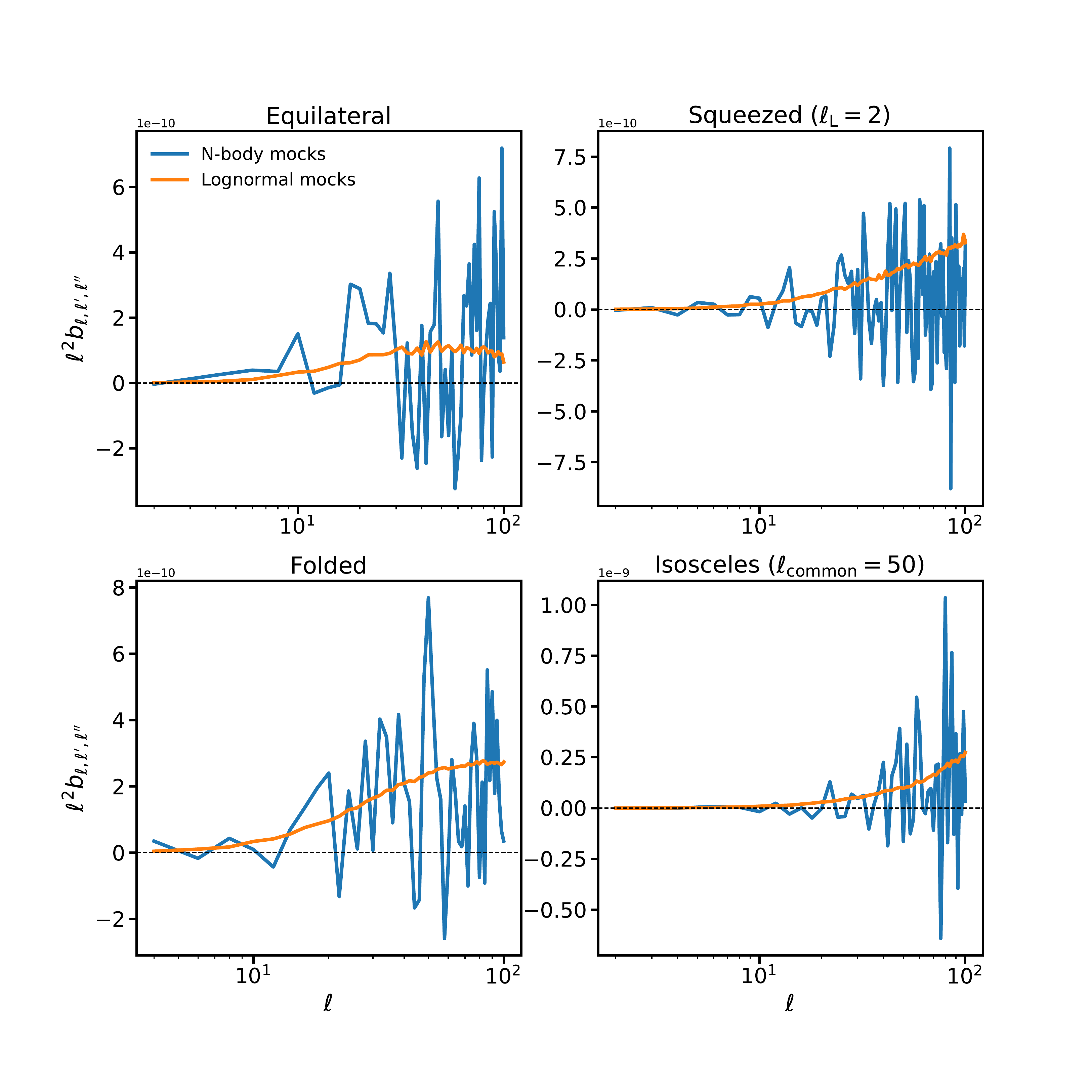}
\caption{Bispectra of our lognormal mocks (orange) and the $N$-body mocks of Ref.~\cite{2017ApJ...850...24T} (blue). We show an equilateral configuration (top left), a squeezed configuration with long wavelength mode $l_{\mathrm{L}} = 2$ (top right), a folded configuration with $l = 2l' = 2l''$ (bottom left), and an isosceles configuration with common side length $l_{\mathrm{common}} = 50$ (bottom right). The bispectra have not been binned in multipole.} 
\label{fig:bispec}
\end{figure}

In Figure~\ref{fig:bispec} we show the reduced bispectrum measured from our lognormal convergence maps and from the $N$-body mocks of Ref.~\cite{2017ApJ...850...24T}. The agreement is reasonable, although the noise in the $N$-body measurement is large. Note that the strength of lognormality was chosen to give a good match to the non-Gaussian part of the power spectrum covariance, so the agreement in the bispectrum is not guaranteed by construction. We also find qualitative agreement with the measurements of Ref.~\cite{2020MNRAS.493.3985M}.

To mitigate the impact of noise in our likelihood we fit the non-Gaussian signal polyspectra with smooth multivariate fifth-order polynomials. To make these low-order polynomials good fits we first bin the polyspectra, as described in Section~\ref{subsec:binning}, in equally spaced bins in $\log l$. We separately fit a univariate polynomial to the ratio of the diagonal elements of the covariance to the Gaussian prediction and a bivariate polynomial to the correlation matrix (excluding the diagonal elements). We fit a trivariate polynomial to the reduced bispectrum before squaring and forming the quantity $\tilde{B}_{l_1 l_2 l_3}$. We take care to include only the non-zero elements of the bispectrum in the fit. 
%

With $T_{ll'}$ and $\tilde{B}_{ll'l''}$ in hand we can now compute the size of the correction to weak lensing likelihoods and posteriors from signal non-Gaussianity, using Equation~\eqref{eq:fullcorrection}.

\subsection{Results}
\label{subsec:results}

\begin{figure}
\centering
\includegraphics[width=0.6\columnwidth]{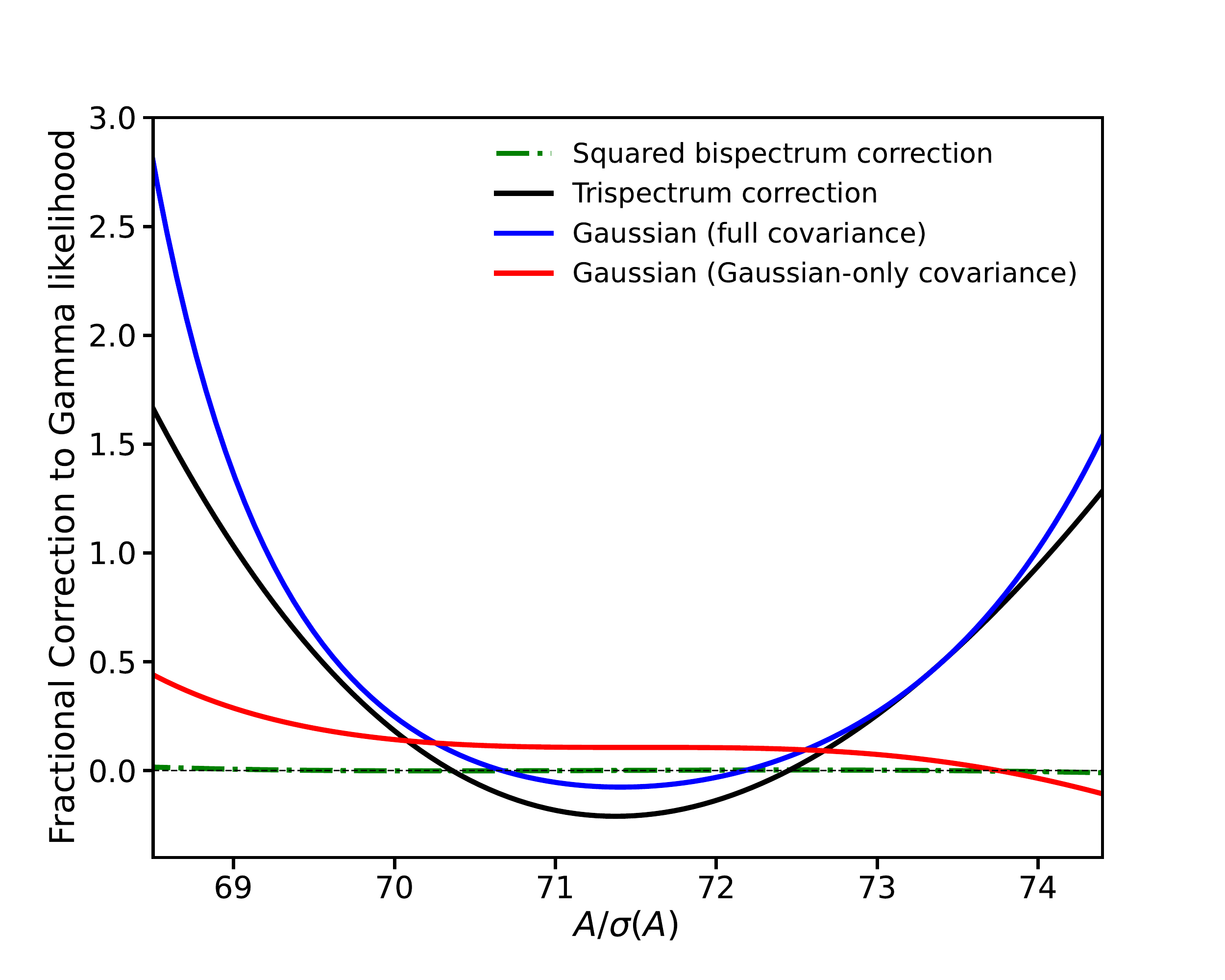}
\caption{Fractional correction to the Gamma likelihood from a non-Gaussian signal. We assume a fixed measurement of the power spectrum $\hat{C}_l$ and look at the correction to the likelihood as function of the model $C_l$. Additionally all values of the model $C_l$ have been scaled relative to $\hat{C}_l$ by an amplitude parameter, $C_l = A\hat{C}_l$. The corrections are shown as a function of $A$ scaled by its Gamma $1\sigma$ uncertainty $\sigma$, i.e.~its signal-to-noise. The non-Gaussian covariance (trispectrum) term (black solid) and squared bispectrum terms (dot-dashed green) are shown separately, along with the fractional difference of a Gaussian with covariance containing only a Gaussian part (red) or additionally the non-Gaussian part (blue).}
\label{fig:likelihood}
\end{figure}

The zero-order likelihood function for fixed $\hat{C}_l$ is proportional to an Inverse-Gamma distribution. In Figure~\ref{fig:likelihood} we show the fractional correction away from this due to signal non-Gaussianity, using $\hat{C}_l$ that have been binned in multipole with $\Delta \log l = 0.1$\footnote{The sensitivity to binning choices is mild, and discussed in Section~\ref{subsec:binning} and Appendix~\ref{app:dlog0p2}.}. In this figure we assume a fixed measured $\hat{C}_l$ and show the correction to the likelihood when the model $C_l$ are uniformly scaled relative to $\hat{C}_l$, i.e.~we take $C_l= A\hat{C}_l$ for an amplitude parameter $A$. The width of the likelihood function is typically $\sigma \equiv 1/\sqrt{\sum \nu/2}$, the zero-order Gamma (or Wishart when $p \neq 1$) uncertainty, so we normalise the horizontal scale in Figure~\ref{fig:likelihood} by this quantity. The best-fit value of $A$ is close to unity, so the actual values plotted on the horizontal axis reflect the $l_{\mathrm{max}}=100$ that we have assumed for this plot. We plot the contributions from the trispectrum and the squared bispectrum separately. We have verified that the `binned' version of our model for the sampling distribution provides a close match to the histogram of measured power spectra from the lognormal mocks across the range of scales contributing to the all the figures in the section, although we note that the mean and covariance are both matched by construction.

Figure~\ref{fig:likelihood} shows that the contribution from the squared bispectrum $\tilde{B}_{ll'l''}$ is sub-dominant to that of the trispectrum in the vicinity of a few $\sigma$ around the best-fit model, which is the range of interest. As expected due to the high mode count contributing to information on $A$ the zero-order distribution is close to Gaussian, as demonstrated by the red curve - note that uniform contributions to the fractional difference will get normalised away when we compute posteriors, as shown shortly. In blue we show the fractional correction of a Gaussian having the correct covariance matrix including the trispectrum contribution. In agreement with the discussion in Section~\ref{subsec:CLT}, this provides an accurate approximation to our distribution, with the main effect being a quadratic correction corresponding to the broadening of parameter error bars.

\begin{figure}
\centering
\includegraphics[width=\columnwidth]{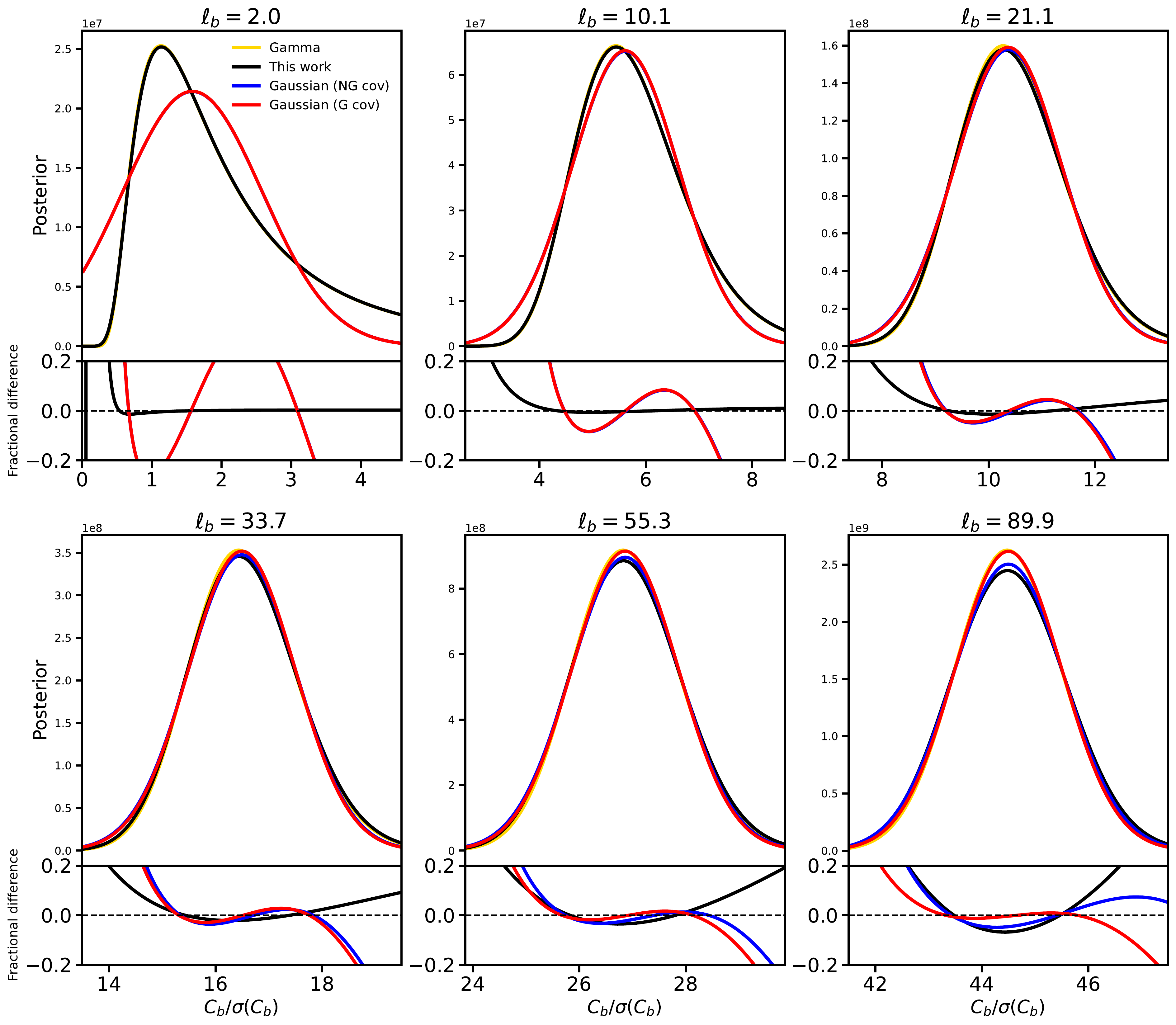}
\caption{Posteriors of individual power spectrum bandpowers, assuming all others are fixed to their measured values, labelled by their weighted central values above each panel. A Jeffreys prior has been assumed for all distributions, and we use scales up to $l_{\mathrm{max}} = 100$. Our model (black) is indistinguishable from the Gamma likelihood (yellow) in the upper panels. We also show a Gaussian with Gaussian covariance (red) and non-Gaussian covariance (blue). Differences between these are indistinguishable in the upper panels, and the red curve is very close to the yellow curve in the lower panels. Corrections are shown as a function of the model band power scaled by its Gamma $1\sigma$ uncertainty $\sigma(C_b)$, i.e.~its signal-to-noise. Our model transitions from Gamma at large angular scales to a Gaussian with non-Gaussian covariance at small angular scales. Below each panel we show the fractional differences with respect to the Gamma likelihood scenario.}
\label{fig:condclpost}
\end{figure}

In practice we are not so much interested in the likelihood as the \emph{posterior}. The easiest posterior to compute from our likelihood is the conditional posterior of a single $C_l$ assuming all others are fixed to their measured values. This is shown in Figure~\ref{fig:condclpost} for $l_{\mathrm{max}} = 100$ and again using binned power spectra with $\Delta \log l = 0.1$, which is the same binning choice used to fit the polyspectra from the simulations\footnote{Note that the binning implementation can differ between that used in the likelihood and that used to fit the polyspectra, but we found using different choices gave negligible effect on the results.}. The bin barycenters are shown above each panel in Figure~\ref{fig:condclpost}. On large angular scales (top left panel), non-Gaussianity in the signal is weak and our distribution matches the Gamma distribution very closely (black and yellow curves are indistinguishable). A Gaussian approximation, in contrast, is poor on these scales. As the angular scale of the multipole bin is decreased, the Gaussian and Gamma approximations match ever more closely, with very little difference between any of the distributions at $l \approx 30$. As the angular scale decreases further, the effects of signal non-Gaussianity start to become important and broaden the distribution through the trispectrum contribution to the covariance matrix. Thus, our distribution starts to deviate from the Gamma and Gaussian-with-Gaussian-covariance distributions (themselves now indistinguishable) visibly at $l \approx 100$. Most of this deviation is captured by a Gaussian with a non-Gaussian covariance (blue curve), with the residual `post-Gaussian' effect being a small amount of skewness in the distribution. In contrast to the Gamma and Gaussian approximations, our distribution is the only model that correctly matches the large-scale and weakly non-linear-scale limits of the true $C_l$ distribution. Similar statements can be made of the marginal distribution of the $C_l$, whose leading-order properties were discussed in Section~\ref{subsubsec:post}.

It is interesting to note from Figure~\ref{fig:condclpost} that the regime where Gamma and Gaussian differ ($l \lesssim 30$ in this case) is distinct from the regime where signal non-Gaussianity starts to become important ($l \gtrsim 100$). That these regimes are well separated has traditionally been the motivation for considering large-scale likelihood non-Gaussianity from the finite number of modes as a separate effect to the small-scale modification of the covariance from signal non-Gaussianity. The distinction is of course redshift dependent; as the redshift of a source bin is lowered, the angular scale at which signal non-Gaussianity becomes important will increase. We therefore expect that our new likelihood expression will be of unique value when these two sources of likelihood non-Gaussianity must be treated simultaneously, i.e.~for low-redshift sources. The precise impact will additionally depend on the parameters of interest. We explore a realistic low-$z_s$ scenario in Appendix~\ref{app:zs6}, finding that skewness corrections to the posterior can become sizeable (although still $<10\%$) for $l_{\mathrm{max}} = 1000$, but are sub-percent once shape noise is included.

%
%

\begin{figure}
\centering
\includegraphics[width=0.6\columnwidth]{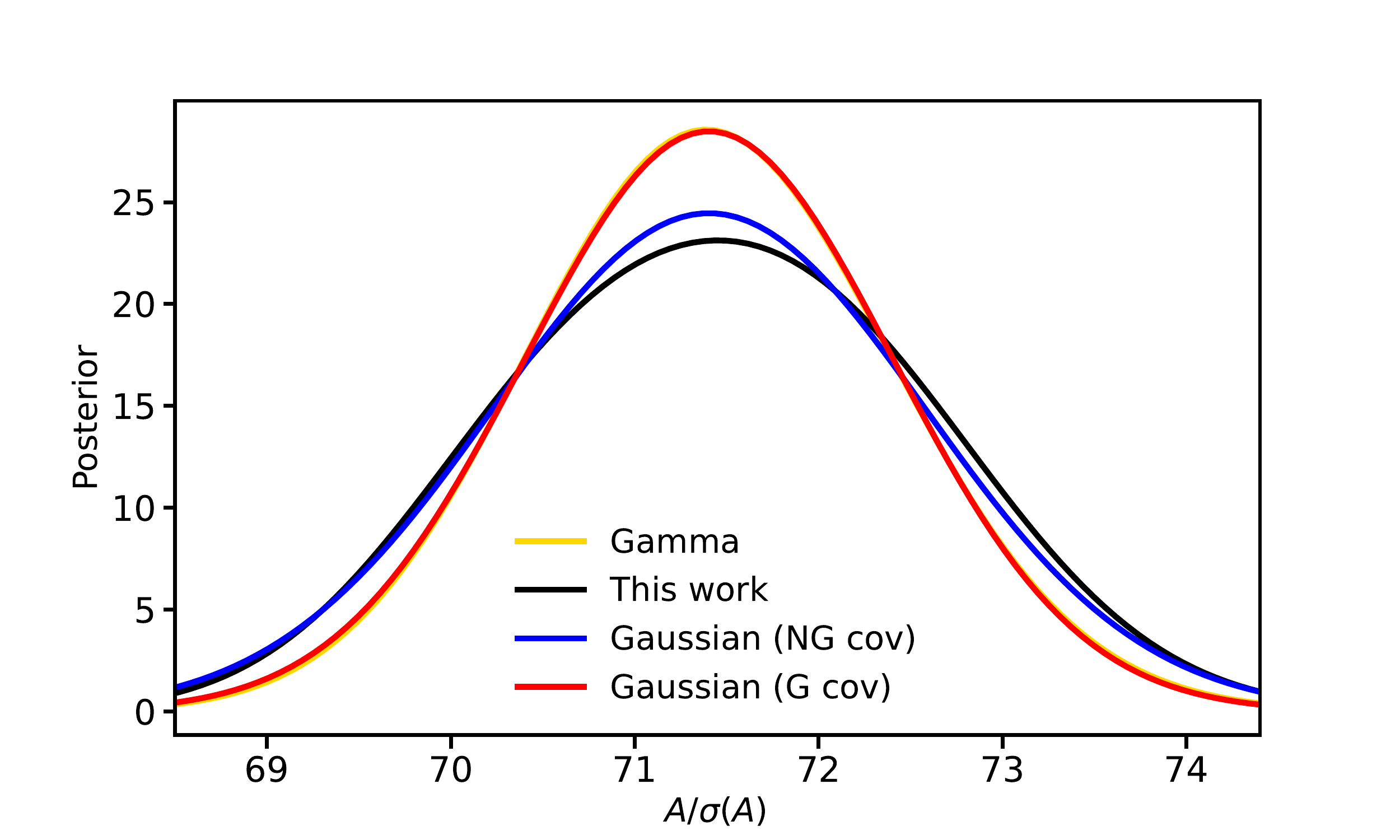}
\caption{Posterior of an amplitude parameter $A$ linearly scaling the model power spectrum, as a function of $A$ scaled by its Gamma $1\sigma$ uncertainty. A Jeffreys prior has been assumed, and we have included scales up to $l_{\mathrm{max}} = 100$. From top to bottom in the centre of the figure we plot the Gamma likelihood (yellow), a Gaussian likelihood with Gaussian covariance matrix (yellow, overlapping with red), a Gaussian likelihood with non-Gaussian covariance matrix (blue), and our model (black).}
\label{fig:amppost}
\end{figure}

It is also straightforward to compute the posterior of an amplitude parameter scaling the signal power spectrum (we again neglect noise for simplicity) as $C_l= A\hat{C}_l$, shown in Figure~\ref{fig:amppost}. Most of the constraining power on the amplitude comes from small scales, and hence the Gamma and Gaussian-with-Gaussian-covariance (yellow and red curves respectively) overlap closely. Our distribution (black curve) correctly captures the significant increase in parameter variance from the trispectrum. As we have seen in Figure~\ref{fig:condclpost}, a Gaussian with appropriately modified covariance (blue curve) captures most of the effect of signal non-Gaussianity, with our distribution imparting a small residual skewness into the posterior.

\begin{figure}
\centering
\includegraphics[width=\columnwidth]{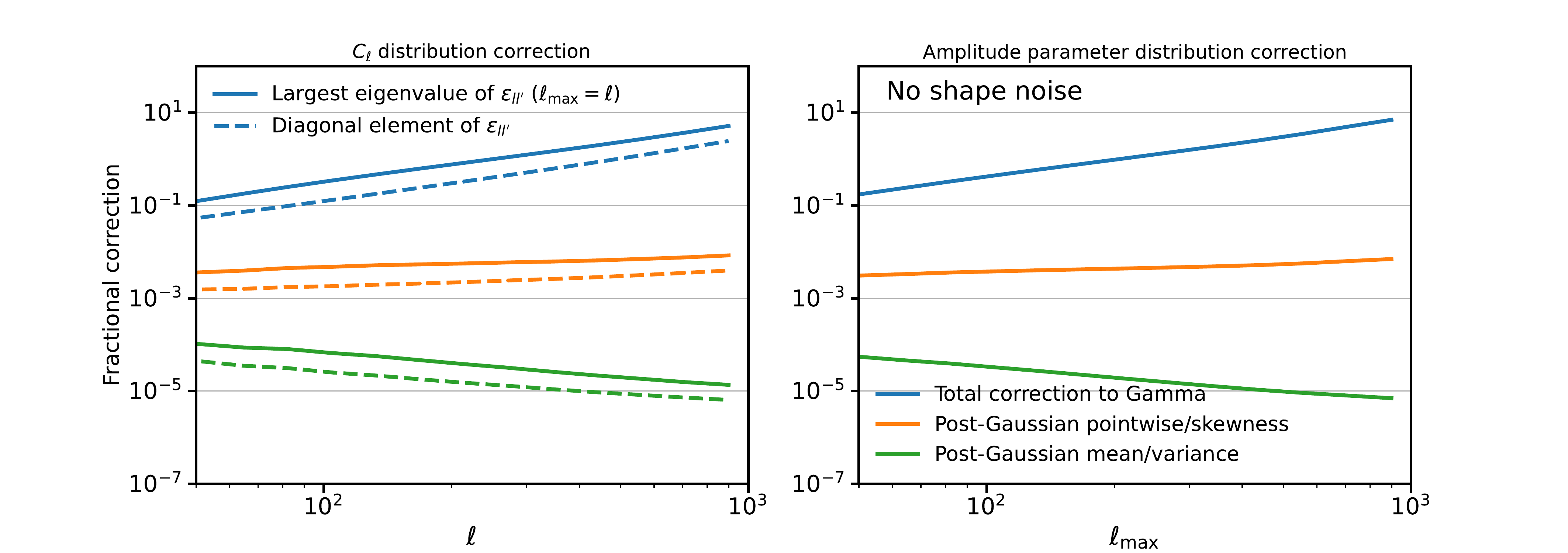}
\caption{Leading-order fractional corrections to the posterior from a non-Gaussian signal as a function of angular wave number, valid in the $l \gg1$ limit. \emph{Left panel}: corrections to the Gamma posterior (conditional or marginal) of the model $C_l$. Shown are the spectral radius (i.e.~largest eigenvalue) of the $\varepsilon_{l l'}$ matrix defined in Equations~\eqref{eq:epsdef} and \eqref{eq:varepsilon_av} (solid blue) and its diagonal elements (dashed blue). These quantities (blue top set of curves) are required to be small for higher-order non-Gaussian terms to be negligible. Leading-order corrections to the Gamma posterior not captured by a Gaussian functional form (`post-Gaussian' effects) are represented by the orange curves (middle set), and sub-dominant corrections to the mean and variance by the green curves (lower set). These are suppressed by factors of $\nu^{-1/2}$ and $\nu^{-1}$ respectively. \emph{Right panel}: Equivalent corrections to the posterior of an amplitude parameter, now showing the cumulative statistic $\langle \varepsilon \rangle $ (see text) up to $l_{\mathrm{max}}$ (blue, top curve) and leading post-Gaussian corrections (orange and green). The interpretation of these curves is the same as in the left panel, i.e.~they represent the total correction to the Gamma posterior from a non-Gaussian signal (blue), and the corrections not accounted for by a Gaussian at leading order in $\nu^{-1/2}$ (orange), and next-to-leading order (green).}
\label{fig:epsmat}
\end{figure}

Since our formalism is based on a perturbative series expansion of the signal distribution, it is important to check that we are working in a regime where higher-order terms can be safely neglected. In Section~\ref{subsubsec:post} we showed that the leading-order correction to the posterior from signal non-Gaussianity is $\langle \varepsilon \rangle$, defined in Equation~\eqref{eq:varepsilon_av}. This quantity is the trispectrum contribution to the $\hat{C}_l$ covariance matrix divided by the Gaussian contribution, $\varepsilon_{l_1 l_2}$, averaged over the scales contributing to the parameter of interest with a $\nu$-dependent weight. It has a simple interpretation that it is equal to the perturbation to the Fisher information from signal non-Gaussianity. We require $\langle \varepsilon \rangle \ll 1$ for our perturbative series to be valid, with the leading-order post-Gaussian (i.e.~beyond that predicted by a Gaussian functional form) skewness of order $\langle \varepsilon \rangle/\sqrt{\sum \nu}$. We have found bispectrum-squared terms to be subdominant, but they are potentially non-negligible when $l \gtrsim 1000$ by the arguments of Section~\ref{subsec:CLT}.

In Figure~\ref{fig:epsmat} we show a few convergence statistics that can be derived from $\varepsilon_{l_1 l_2}$. The left panel shows quantities relevant to the posterior of a single $C_l$, in which case $\langle \varepsilon \rangle$ is equivalent to $\varepsilon_{ll}$, the diagonal elements of the matrix $\varepsilon_{l_1 l_2}$. We also show the largest eigenvalue (i.e.~the spectral radius) of this matrix. A spectral radius much less than unity is sufficient to ensure a valid perturbative expansion of the inverse covariance matrix around its Gaussian term, and must hold if our expression is to give a valid approximation to the pairwise distributions of $C_l$ with different $l$.

As the blue curves in Figure~\ref{fig:epsmat} show, the fractional correction from signal non-Gaussianity is much less than unity out to $l \approx 100$, which justifies our use of this maximum scale in the posteriors computed in this section and explains why the correction appears so small in Figure~\ref{fig:condclpost}. The leading-order corrections not captured by a Gaussian functional form are given by the orange curves and are never more than $10^{-2}$ even out to $l \approx 1000$, i.e.~sub-percent across all scales of interest for upcoming lensing surveys, an important result. Beyond $l \approx 100$ our perturbative series breaks down, but the quantity $\langle \varepsilon \rangle/\sqrt{\sum \nu}$ is still indicative of the relevance of residual non-Gaussianity in the likelihood, so it is reasonable to draw useful conclusions from Figure~\ref{fig:epsmat} even beyond the regime where our likelihood expression is formally valid. We note that $\langle \varepsilon \rangle/\sqrt{\sum \nu}$ is the leading-order correction to the dimensionless skewness of the measured power spectrum from signal non-Gaussianity, valid at all (sufficiently large) $l$, so it is reasonable to expect that this quantity also quantifies non-Gaussian corrections to the likelihood function. Corrections to the posterior mean and variance not captured by a Gaussian are even more subdominant, suppressed by another factor of $\sqrt{\nu}$.

The right-hand panel of Figure~\ref{fig:epsmat} shows $\langle \varepsilon \rangle$ for an amplitude parameter scaling the model power spectrum. The corrections are comparable in size with those of the individual $C_l$, with $l_{\mathrm{max}} = 100$ just about in the regime where our expression is valid. Non-Gaussian effects are still sub-percent even out to $l_{\mathrm{max}} = 1000$. In Appendix~\ref{app:dlog0p2} we study the (mild) impact of choosing a different binning scheme for the power spectrum bandpowers, shown in Figure~\ref{fig:epsmat_0p2}.

\begin{figure}
\centering
\includegraphics[width=\columnwidth]{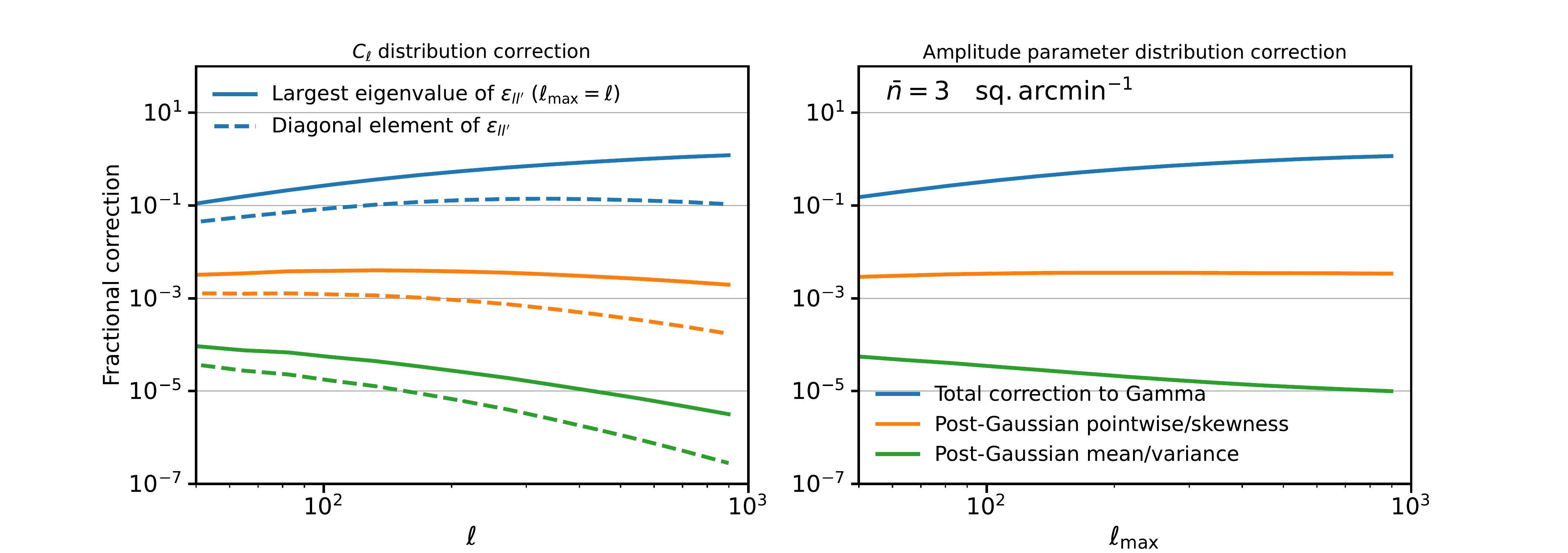}
\caption{Same as Figure~\ref{fig:epsmat} but with shape noise corresponding to $\bar{n} = 3 \, \mathrm{sq. \, arcmin}^{-1}$.}
\label{fig:epsmat_n3}
\end{figure}

Taken together, Figures~\ref{fig:condclpost} and~\ref{fig:epsmat} give the following picture for how the likelihood changes with angular multipole. On the largest scales a Gamma distribution is accurate as the signal is close to Gaussian. By $l \approx 30$ the Gamma distribution has Gaussianized, with corrections being percent-level or small in the central $2\sigma$ region (albeit larger in the tails). As $l \approx 100$ is approached the non-Gaussianity in the signal starts to become important, and a Gaussian with appropriately broadened covariance is a good approximation. Corrections to the Gamma distribution not captured by the Gaussian approximation are sub-percent across all angular scales, but can be accounted for by using our expression Equation~\eqref{eq:fullcorrection} on large scales and smoothly matching it to a Gaussian around $l \lesssim 100$.

One question we wish to answer in this work is whether the strength of signal non-Gaussianity grows faster than the CLT is able to Gaussianize the power spectrum distribution. From Figure~\ref{fig:epsmat} it appears that the struggle between these two competing effects is delicately balanced, with a very slow growth in the post-Gaussian correction from signal non-Gaussianity (orange curve) visible as $l_{\mathrm{max}}$ increases. One might wonder whether this correction becomes non-negligible at very high $l_{\mathrm{max}}$. At some point as $l_{\mathrm{max}}$ is increased the effect of the noise will become important, which suppresses $\varepsilon$ by boosting $C_l$ while not adding any non-Gaussianity (we assume the noise is Gaussian here). In Figure~\ref{fig:epsmat_n3} we show the size of post-Gaussian effects (i.e.~corrections to the Gamma distribution not captured by a Gaussian likelihood) on the posterior with shape noise given by $\sigma_\gamma^2 = \sigma_e^2/\bar{n}$, i.e.~the model power spectrum is $C_l = AS_{0,l} + N_l$. We assume a per-component ellipticity r.m.s. of $\sigma_e = 0.21$, and set the source number density to $\bar{n} = 3 \, \mathrm{arcmin}^{-2}$, as might be expected in a single tomographic bin in a Euclid-like survey with 10 redshift bins~\cite{2020A&A...642A.191E}. The Gaussianizing effects of noise start to become important at $l \gtrsim 100$, and ensure that post-Gaussian effects in the likelihood are never more than sub-percent. For $\bar{n} =  30 \, \mathrm{arcmin}^{-2}$ this turn-over happens around $l_{\mathrm{max}} \approx 1000$, suggesting that post-Gaussian effects are safely negligible on small scales for Stage-IV surveys.

As discussed above, it will likely be the case that post-Gaussian effects are most relevant when the largest angular scales in the survey are mildly non-linear, i.e.~for low-redshift tomographic bins. These carry little statistical weight in a Euclid-like survey where most of the sources are around $z\approx 1$, but could be relevant in other surveys. In Appendix~\ref{app:zs6} we explore the corrections for the lowest source redshift bin of a 10-bin lensing survey having intrinsic source redshifts comparable to that of a Euclid-like survey, showing that corrections are sub-percent once the effects of shape noise are included. This is partly because the amplitude of the shear power spectrum is suppressed for nearby sources due to the lensing geometry, so it may be that effects are larger for projected galaxy clustering. The quantity $\langle \varepsilon \rangle /\sqrt{\sum \nu}$, or alternatively the dimensionless skewness of the measured power spectrum, provides an easy-to-compute diagnostic of whether one needs to worry about non-Gaussianity in the likelihood when low-redshift bins are important. If the effects are large, one will also need to account for effects of the mask in a full pseudo-$C_l$ analysis to build an accurate likelihood model, as in Ref.~\cite{2008PhRvD..77j3013H, 2009MNRAS.400..219B, 2020MNRAS.491.3165U, 2021MNRAS.507.1072D}, or resort to map-level or likelihood-free methods.


\section{Conclusions}
\label{sec:conc}

We have computed, for the first time, the leading-order correction to the distribution of angular power spectrum estimates of a projected field from non-Gaussianity in the underlying signal. Our expressions are given in Equation~\eqref{eq:fullcorrection} in the case of a single map, and by the sum of Equation~\eqref{eq:finalpCl_tri_pne1} and Equation~\eqref{eq:finalpCl_bisq_pne1} in the case of multiple correlated maps. These expressions correct the Wishart distribution for the matrix of power spectra, and consist of a term proportional to the signal trispectrum and a term proportional to the square of its bispectrum. The expression can easily be generalised to include non-Gaussian noise.

Our expressions are correctly normalised, and have the correct mean, covariance, and three-point function. We have shown that, in the limit of a large number of degrees of freedom and sufficiently well behaved non-Gaussianity, the main effect of signal non-Gaussianity is to correct the covariance matrix of the power spectra by the standard trispectrum term. The residual `post-Gaussian' effect of non-Gaussianity is to impart some skewness to the likelihood, on top of that present due to the residual effects of finite degrees of freedom. On small scales, the size of the correction to the posterior of a parameter is given approximately by the fractional correction to the Fisher information from the trispectrum, scaled by the inverse square-root of the total degrees of freedom that contribute. For a given $C_l$ and $l \gg 1$ the relevant quantity is $\Delta_l \equiv \sqrt{l/2} T_{ll}/C_{l}^2$, where $T_{ll}$ is the trispectrum contribution to the variance. When $\Delta_l \ll 1$, corrections to the likelihood of the $C_l$ that are not captured by a Gaussian with appropriately modified covariance are suppressed.

We have provided diagnostic statistics, given in Equation~\eqref{eq:epsdef} and Equation~\eqref{eq:varepsilon_av} that quantify corrections to likelihoods and posteriors. These can be computed straightforwardly from a model for the non-Gaussian part of the covariance matrix. Such a model is already a requirement when using a likelihood for two-point statistics, so these statistics should be easy to compute for a given observational setup.

We quantified corrections to the likelihood function using mock non-Gaussian weak lensing maps created assuming lognormal statistics, with input power spectra and noise levels representative of an upcoming Stage-IV survey. The bispectrum-squared term is subdominant to the correction from the trispectrum at least to $l\approx 100$, but we do not rule out stronger effects on small angular scales. Corrections to the likelihood not captured by a Gaussian are strongly suppressed on small scales in all our tests, which provides evidence that a Gaussian likelihood is sufficient on small scales even in the presence of realistic signal non-Gaussianity. 

On large angular scales our likelihood correctly tends to a Gamma/Wishart distribution. We have found that the angular scales where the Gamma transitions to a Gaussian are sufficiently well separated from those where the trispectrum becomes important that it is sufficient to approximate the functional form of the likelihood as Gaussian when incorporating signal non-Gaussianity. This is the approach that has been taken traditionally, albeit without rigorous justification. This property remains true for the lowest redshift bin of a 10-bin Euclid-like lensing survey, but may break down for galaxy clustering power spectra at low redshifts. The likelihood function we have derived is unique amongst commonly used likelihoods in that it has the correct behaviour on both large and small angular scales, being non-perturbative in the degrees of freedom.

Our model relies on the signal non-Gaussianity being weak in a certain, well-defined, sense. We have taken care to identify the regime where this approximation is valid, but have argued that requiring $\Delta_l \ll 1$ is a good diagnostic for the post-Gaussian effects even beyond the limiting scale where our expansion is valid. Use of our likelihood in a practical situation will require ensuring that $(2l+1)\Delta_l \ll 1$, or more generally that the corrections from signal non-Gaussianity are small in the likelihood or posterior of interest. This can also be checked by ensuring that non-Gaussianity in the sampling distribution of the relevant combinations of the \emph{measured} power spectrum is small, which can be checked with simulations (which are in any case necessary to validate or produce covariance matrix estimates).

We have elucidated how the Central Limit Theorem competes with signal non-Gaussianity in suppressing non-Gaussian corrections to the likelihood. The balance is surprisingly delicate, but our conclusion is that our Universe is not non-Gaussian enough on the relevant scales that post-Gaussian terms in the likelihood are necessary in the high-$l$ regime -- the Central Limit Theorem wins. This conclusion is strengthened once Gaussian noise (e.g.~from galaxy intrinsic shapes or shot noise) is present in the data. This essentially explains why previous works such as Ref.~\cite{2019PhRvD.100b3519T, 2020MNRAS.499.2977L} have found a Gaussian likelihood to be sufficient for $\Lambda$CDM parameter inference in the presence of realistic signal non-Gaussianity. Our findings are also consistent with the results of Ref.~\cite{2020PhRvD.102j3507D}, who found that by far the dominant source of non-Gaussianity in the power spectrum likelihood comes from large angular scales where signal non-Gaussianity is negligible. Our formalism has allowed us to cleanly separate the effects of non-Gaussianity in the signal with that from the quadratic nature of the power spectrum estimator, improving upon previous numerical works that have been unable to distinguish between these distinct effects.

An important limitation of our likelihood is that it is only valid for full-sky power spectrum estimates. For the standard pseudo-$C_l$ estimator even the zero-order Gaussian-field likelihood is difficult to compute~\cite{2020MNRAS.491.3165U}, and accounting for the mode coupling entailed by a survey mask in the non-Gaussian case seems intractable at first sight. The main effect of the mask is to reduce the number of degrees of freedom, increasing parameter variances. As the size of the survey is reduced, we can expect that at some point the largest observable scale is non-linear, in which case a suitably modified version of our likelihood might be a good place from which to build an approximation. The effects of non-Gaussianity from the signal in the case of small surveys may be behind the fairly large posterior corrections found in Ref.~\cite{2009A&A...504..689H}. In the case that the full-sky power spectra are estimated from the data directly (as in quadratic maximum likelihood approaches) our distribution is likely to be more accurate, and bears some resemblance to the approximations developed in Ref.~\cite{2017JCAP...12..009S}. For small survey cuts the mode coupling induced by the mask may only be relevant in the regime where the signal is linear, in which case an $f_{\mathrm{sky}}$ correction to the number of degrees of freedom may be sufficient to incorporate the survey mask into our likelihood. However, it may be the case that a map-level likelihood such as that used in cosmic microwave background analyses~\cite{2004PhRvD..70h3511W}, a likelihood-free approach~\cite{2019MNRAS.488.4440A}, or suitable data transformations~\cite{2019MNRAS.486..951W} may be the way forward for modelling these effects in the distribution of projected fields.

While this work has studied the power spectrum likelihood, most weak lensing surveys up to date have used the shear correlation functions to derive their cosmological constraints. An analytic expression for the correlation function likelihood is difficult to write down even in the case of linear shear fields, but may be inferred numerically by sampling power spectra and subsequently transforming to real space. We defer an analytic study of this to a future work, but note that the results of the numerical studies Ref.~\cite{2020MNRAS.499.2977L} and Ref.~\cite{2019PhRvD.100b3519T} suggest that non-Gaussian corrections are likely negligible for $\Lambda$CDM models.
 
This work represents the first rigorous analytic study of the use of a Gaussian approximation for likelihood function of the power spectrum of non-Gaussian fields. While residual non-Gaussian effects appear small on small angular scales, we recommend that this be checked on a survey-by-survey basis using the diagnostic statistics defined in Equations~\eqref{eq:epsdef} and~\eqref{eq:varepsilon_av}. With the huge increase in statistical constraining power offered by near-future wide imaging surveys, it will be increasingly important to ensure that no biases arise from an incorrect statistical description of the data. Since we have found no evidence that signal non-Gaussianity imparts significant non-Gaussianity into the likelihood beyond that coming from the finite degrees of freedom, our recommendation for upcoming Stage-IV lensing surveys is to model the power spectrum likelihood as Gaussian (with the correct non-linear covariance matrix) on small scales $l \gtrsim 30$. On large scales in the case that mode coupling from the mask is irrelevant the Gamma (or Wishart in the case of multiple maps) distribution should be used, and when mode coupling is relevant one should resort to map-level, likelihood-free, or existing likelihood approximations. The likelihood we have derived in this work may be used as a bridge between these two regimes.

\begin{acknowledgments}

We thank Karim Benabed, Michael Brown, Alan Heavens, and Robin Upham for useful conversations. We thank Eric Tittley for his maintenance of and help with cuillin, the IfA's computing cluster (~\url{http://cuillin.roe.ac.uk}) partially funded by the Science and Technology Facilities Council (STFC) and the European Research Council. Some of the results in this paper have been derived using the HEALPix~\cite{2005ApJ...622..759G} package. AH and ANT acknowledge the support of a UK STFC Consolidated Grant.
\end{acknowledgments}



\appendix

\section{Power spectrum cumulants and the Edgeworth expansion}
\label{app:Edge}

When considering how non-Gaussianity in a signal affects the distribution of the power spectrum, it is useful to start by studying the first few cumulants. In this section we will keep the assumption of Gaussian noise, and specialise to a single redshift bin ($p=1$) for simplicity.

Firstly, let us compute the covariance of the power spectrum estimates. Non-Gaussianity contributes via the trispectrum of the signal as
\begin{equation}
\langle \Delta \hat{C}_l \Delta \hat{C}_{l'} \rangle = \frac{2}{\nu}C_l^2 \delta_{l l'} + T_{ll'},
\label{eq:NGcov}
\end{equation}
where $T_{ll'}$ is an $m$-averaged version of the signal trispectrum; explicit forms for this are given in Ref.~\cite{2002PhRvD..66f3008O}. We remind the reader here that $C_l$ is the sum of signal and noise power. The trispectrum term $T_{ll'}$ is in general difficult to compute, with current strategies in weak lensing focused around halo models~\cite{2002PhR...372....1C, 2017MNRAS.470.2100K} or $N$-body simulations~\cite{2012MNRAS.426.1262H, 2013PhRvD..87l3538S}.

Now consider the three-point function of $\hat{C}_l$. This consists of terms having six $a_{lm}$ factors. We can compute this from the definition Equation~\eqref{eq:Chatdef}, using Wick's theorem to pick out the various contributions. This gives
\begin{equation}
    \langle \Delta \hat{C}_{l_1}\Delta \hat{C}_{l_2}\Delta \hat{C}_{l_3}\rangle =  8\frac{\delta_{l_1l_2}\delta_{l_2l_3}}{\nu_1^2}C_{l_1}^3
     + 4\tilde{B}_{l_1l_2l_3} + 2\tilde{S}_{l_1 l_2 l_3} + 4[3]\frac{\delta_{l_1l_2}}{\nu_1}C_{l_1}T_{l_1 l_3} + P_{l_1 l_2 l_3},
    \label{eq:skewofCl}
\end{equation}
where the notation $[3]$ denotes the $3$ terms resulting from symmetrization over wave numbers in terms to the right, $P_{l_1 l_2 l_3}$ is an $m$-averaged connected signal six-point function (`pentaspectrum') and the bispectrum-squared quantities $\tilde{B}$ and $\tilde{S}$ are defined as
\begin{align}
    \tilde{B}_{l_1l_2l_3} &\equiv \frac{1}{\nu_1 \nu_2 \nu_3} \sum_{m_1,m_2,m_3} \langle s_{l_1m_1}s_{l_2m_2}s_{l_3m_3}\rangle \langle s_{l_1m_1}^{*}s_{l_2m_2}^{*}s_{l_3m_3}^{*}\rangle, \label{eq:btilde} \\
    \tilde{S}_{l_1l_2l_3} &\equiv \frac{1}{\nu_1 \nu_2 \nu_3} \sum_{m_1,m_2,m_3} \langle s_{l_1m_1}s_{l_1m_1}^*s_{l_2m_2}\rangle \langle s_{l_2m_2}^{*}s_{l_3m_3}s_{l_3m_3}^{*}\rangle + \mathrm{sym.}\label{eq:stilde}
\end{align}
By statistical isotropy the signal bispectra must be invariant under rotations, which constrains the terms in angle brackets in Equations~\eqref{eq:btilde} and~\eqref{eq:stilde} to be proportional to $3j$-symbols that carry all the $m$-dependence~\cite{2000ApJ...534..533C}. We will further impose that the signal statistics are invariant under a parity inversion, which introduces a second $3j$-symbol imposing that all the wave numbers sum to an even integer. Explicitly we have
\begin{equation}
\langle a_{l_1m_1}a_{l_2m_2}a_{l_3m_3} \rangle = \sqrt{\frac{\nu_1\nu_2\nu_3}{4\pi}} \ThreeJSymbol{l_1}{0}{l_1}{0}{l_3}{0} \ThreeJSymbol{l_1}{m_1}{l_1}{m_2}{l_3}{m_3}b_{l_1l_2l_3},
\end{equation}
where $b_{l_1 l_2 l_3}$ is the reduced bispectrum. Substituting this expression into Equations~\eqref{eq:btilde} and~\eqref{eq:stilde}, we see that $\tilde{S}$ is proportional to terms like
\begin{equation}
    \sum_{m_1} (-1)^{m_1} \ThreeJSymbol{l_1}{m_1}{l_1}{-m_1}{l_3}{m_3} = \delta_{l_3 0}\delta_{m_3 0}(-1)^{l_1}\sqrt{2l_1+1},
\end{equation}
where the identity follows from the orthogonality of the $3j$-symbols. Since the signal $l=0$ mode is unobservable, the overall contribution from terms of this form is zero, and hence $\tilde{S} = 0$. For $\tilde{B}$ we have
\begin{equation}
    \tilde{B}_{l_1l_2l_3} = \frac{1}{4\pi} \ThreeJSymbol{l_1}{0}{l_2}{0}{l_3}{0}^2  b_{l_1l_2l_3}^2.
\end{equation}
As mentioned above, the $3j$-symbol imposes that $l_1 + l_2 + l_3$ is an even integer, as well as the triangle conditions $\lvert l_1 - l_2 \rvert \leq l_3 \leq l_1+l_2$.

The three-point function of power spectrum estimates thus receives contributions from squared bispectra, products of trispectra and power spectra, and the connected six-point function. Similarly, the connected four-point function will receive contributions from squared trispectra, squared bispectra multiplied by power spectra, bispectra multiplied by the connected five-point function, pentaspectra multiplied by power spectra, and the connected eight-point function.

We want to keep our model for non-Gaussianity as general as possible, so in principle we would like to keep track of the whole cumulant hierarchy of the power spectrum estimates. This is clearly going to be computationally unfeasible however, so we will hereafter assume that the signal non-Gaussianity is \emph{weak}.

\subsection{Weak non-Gaussianity}
\label{app:weakNG}

We will assume that non-Gaussianity in the signal is weak in the sense that each term in its cumulant hierarchy is successively smaller than the last. Introducing the order-counting parameter $\lambda$, we will assume that $C_l \sim \mathcal{O}(\lambda)$, $b_{l_1l_2l_3} \sim \mathcal{O}(\lambda^2)$, $T_{l_1 l_2} \sim \mathcal{O}(\lambda^3)$, etc., with $\lambda \ll 1$. This is justified on sufficiently large scales where non-linearity in the underlying field is weak, and perturbation theory is accurate. We will precisely define the regime in which this perturbative approach is valid later on.

In this scenario, bispectrum-squared terms like $\tilde{B}_{l_1 l_2 l_3}$ are the same perturbative order as power-trispectrum cross terms like $C_{l_1} T_{l_1 l_3}$, which can also be seen by expanding the fields to second order in the linear signal; the leading-order (tree level) bispectrum features two factors of the linear power spectrum~\cite{2002PhR...367....1B}, meaning the squared bispectrum contains four such factors. The trispectrum contains three power spectra at leading order, and hence $C_{l_1} T_{l_1 l_3}$ also contains four factors of the linear power spectrum. The connected six-point function is suppressed as $P_{l_1 l_2 l_3} \sim \mathcal{O}(\lambda^5)$. We will retain leading-order `non-Gamma' terms throughout. The 3-pt function of $\hat{C}_l$ is thus Equation~\eqref{eq:skewofCl} with $\tilde{S}=0$ and the final $P_{l_1 l_2 l_3}$ term dropped.

The \emph{dimensionless} 1-pt skewness of $\hat{C}_l$ can be derived from Equation~\eqref{eq:skewofCl} as
\begin{align}
\frac{\langle \Delta \hat{C}_l^3 \rangle}{\langle \Delta C_l^2 \rangle^{3/2}} &= \frac{8C_l^3/\nu^2 + 4\tilde{B}_{lll} + 12C_lT_{ll}/\nu}{\left(2C_l^2/\nu + T_{ll}\right)^{3/2}} \nonumber \\
&= \sqrt{\frac{8}{\nu}} + \sqrt{2\nu^3}\frac{\tilde{B}_{lll}}{C_l^3} + \frac{3\sqrt{2\nu}}{2}\frac{T_{ll}}{C_l^2} + \mathcal{O}(\lambda^2).
\label{eq:dimlessskewCl}
\end{align}
It is important to note here that we could consistently neglect the $\mathcal{O}(\lambda^5)$ term $P_{l_1 l_2 l_3}$ in both the 3-pt function and its dimensionless version, since the variance itself is $\mathcal{O}(\lambda^2)$ meaning that connected 6-pt terms contribute at $\mathcal{O}(\lambda^2)$ in the dimensionless skew. This is in contrast to the connected four-point function of the power spectrum estimates. The 1-pt excess kurtosis for example is
%
%
\begin{equation}
\langle \Delta \hat{C}_{l}^4 \rangle_c = 48\frac{C_l^4}{\nu^3} + 96\frac{C_l \tilde{B}_{lll}}{\nu} + 144\frac{C_l^2 T_{ll}}{\nu^2},
\end{equation}
where we have kept terms at $\mathcal{O}(\lambda^5)$. Working to a consistent $\mathcal{O}(\lambda^4)$ order would mean dropping all but the first term from this expression. In contrast, the \emph{dimensionless} kurtosis of $\hat{C}_l$ is
\begin{align}
\frac{\langle \Delta \hat{C}_{l}^4 \rangle_c}{\langle \Delta \hat{C}_l^2 \rangle^2} &= \frac{48C_l^4/\nu^3 + 96C_l \tilde{B}_{lll}/\nu + 144C_l^2 T_{ll}/\nu^2}{\left(2C_l^2/\nu + T_{ll}\right)^{2}} \nonumber \\
&= \frac{12}{\nu} + 24\nu \frac{\tilde{B}_{lll}}{C_l^3} + 24\frac{T_{ll}}{C_l^2} + \mathcal{O}(\lambda^2).
\end{align}
Therefore, while the raw non-Gaussian cumulant hierarchy of the power spectrum estimates can be truncated consistently in perturbation theory, \emph{all} of the dimensionless cumulants, i.e.~the cumulants of a normalised $\hat{C}_l$ estimator with unit variance, receive corrections from a non-Gaussian signal. The cumulant hierarchy of the normalised estimator cannot therefore be truncated consistently, since terms of order $\mathcal{O}(\lambda)$ always contribute.

The higher-order non-Gaussian cumulants of a normalised $\hat{C}_l$ estimator thus do not get successively smaller, in contrast to those of the signal. This observation has important consequences, since successively smaller cumulants are a requirement for the Edgeworth expansion to converge. One might be tempted to speculate that a good way of incorporating the effects of a non-Gaussian signal into the $\hat{C}_l$ distribution would be to treat the non-Gaussianity as weak and then Edgeworth expand around a Gamma distribution, but the argument above tells us this will not work; we have to instead expand the \emph{signal} distribution around a Gaussian. Attempts to use an Edgeworth expansion directly on the distribution of $\hat{C}_l$ (or a decorrelated version of it, as in Ref.~\cite{2020MNRAS.499.2977L}) in order to incorporate signal non-Gaussianity should be treated with caution.

\section{Derivation of the tomographic redshift bin likelihood}
\label{app:pne1}

In this Section we derive the correction to the likelihood function of tomographic redshift bin power spectra. It is straightforward to write down the correct likelihood for full-sky spectra when the signal and noise are both Gaussian - this is the Wishart distribution for positive-definite matrices. Similarly, it is straightforward to write down a Gaussian approximation to the Wishart distribution. Here we present the derivation for the case of multiple redshift bins.

Our strategy will be to substitute the Edgeworth-expanded signal characteristic function Equation~\eqref{eq:Edge} into the mode coupling integral Equation~\eqref{eq:char_1} and perform the integrations to derive the characteristic function of the power spectrum. We will then inverse Fourier-transform this to get the distribution of the power spectrum.

We will compute the trispectrum and bispectrum-squared corrections separately, but start with the simple case of Gaussian signal and noise.

\subsection{Gaussian signal, Gaussian noise}
\label{subapp:GsGn}

As a warm-up, consider the situation where both signal and noise are Gaussian. The former should be a good approximation on large scales where the signal is a linearly-evolved version of the (close to) Gaussian initial conditions, and the latter is practically guaranteed for maps made by compression from large numbers of underlying tracers (e.g.~shear maps constructed from averaged galaxy shapes, or density maps made by counting many galaxies).

We assume throughout that the underlying signal and noise obey statistical isotropy, such that modes with different wave vectors are uncorrelated. When the signal and noise are both Gaussian, the covariance of different modes captures all the statistical information of the field, implying that modes with different wave vectors are not only uncorrelated but statistically independent. The characteristic functions of signal and noise thus factorize, and since the Fourier transform of a Gaussian is also a Gaussian we have
\begin{align}
    \phi_s(\{\mathbf{k}_{lm}\}) &= e^{-\frac{1}{2} \sum_{lm} \mathbf{k}_{lm}^\dagger \mathbfss{S}_{l} \mathbf{k}_{lm}}, \label{eq:charG_pne1}\\
    \phi_n(\{\mathbf{k}_{lm}\}) &= e^{-\frac{1}{2} \sum_{lm} \mathbf{k}_{lm}^\dagger \mathbfss{N}_{l} \mathbf{k}_{lm}},
\end{align}
where $\mathbfss{S}_{l}$ and $\mathbfss{N}_{l}$ are the signal and noise covariance matrices.

Equation~\eqref{eq:char_1} factorizes in each $l$ and $m$, with each integral a Gaussian integral. Making these substitutions immediately gives
\begin{equation}
    \phi_{\{\nu \hat{\mathbfss{C}}_l\}}(\{\mathbfss{J}_l\}) = \prod_{l=l_{\mathrm{min}}}^{l_{\mathrm{max}}} \lvert \mathbfss{I} + 2i\mathbfss{J}_l\mathbfss{C}_l \rvert^{-\frac{\nu}{2}},
\end{equation}
where $\mathbfss{C}_l$ is the total (signal plus noise) covariance matrix of the map. We recognise this as a product of characteristic functions of Wishart matrices.

The joint distribution of the power spectra is thus
\begin{equation}
    p(\{\nu \hat{\mathbfss{C}}_l\}) =  \prod_{l=l_{\mathrm{min}}}^{l_{\mathrm{max}}} 2^{\frac{p(p-1)}{2}}\int \frac{\mathrm{d} \mathbfss{J}_l }{(2\pi)^{n}} e^{i \mathrm{Tr}(\nu \mathbfss{J}_l \hat{\mathbfss{C}}_l)} \lvert \mathbfss{I} + 2i\mathbfss{J}_l\mathbfss{C}_l  \rvert^{-\frac{\nu}{2}},
    \label{eq:Wishart_char_pne1}
\end{equation}
i.e.~each $\hat{\mathbfss{C}}_l$ is independent.

The integral in Equation~\eqref{eq:Wishart_char_pne1} is almost the standard integral that defines the multivariate Gamma function, with the exception that it is over real symmetric matrices rather than positive definite matrices. Now, let $\mathbfss{T} \equiv 2\mathbfss{C}_l^{\frac{1}{2}}\mathbfss{J}_l\mathbfss{C}_l^{\frac{1}{2}}$ and change integration variables from $\mathbfss{J}_l$ to $\mathbfss{T}$. We'll assume the total covariance matrix is positive definite. The Jacobian of this transformation is
\begin{equation}
    \mathrm{d} \mathbfss{J}_l = 2^{-\frac{p(p+1)}{2}} \left \vert \mathbfss{C}_l \right \vert^{-\frac{(p+1)}{2}} \mathrm{d}\mathbfss{T},
\end{equation}
which follows from a standard result. Each term in the product is therefore
\begin{equation}
    \frac{2^{-p} \left \vert \mathbfss{C}_l \right \vert^{-\frac{(p+1)}{2}}}{(2\pi)^{\frac{p(p+1)}{2}}} \int \mathrm{d}\mathbfss{T} \; e^{\frac{i}{2}\mathrm{Tr}(\boldsymbol{\Omega} \mathbfss{T})} \lvert \mathbfss{I} + i\mathbfss{T} \rvert^{-\frac{\nu}{2}}
\end{equation}
where $\boldsymbol{\Omega} \equiv \nu\mathbfss{C}_l^{-\frac{1}{2}}\hat{\mathbfss{C}}_l \mathbfss{C}_l^{-\frac{1}{2}} $ is symmetric and the integration is over all real symmetric $p \times p$ matrices $\mathbfss{T}$. Our task is therefore to compute the integral
\begin{equation}
   P_{p,\nu}(\boldsymbol{\Omega} ) \equiv  \int \mathrm{d}\mathbfss{T} \; e^{\frac{i}{2}\mathrm{Tr}(\boldsymbol{\Omega} \mathbfss{T})} \lvert \mathbfss{I} + i\mathbfss{T} \rvert^{-\frac{\nu}{2}}.
   \label{eq:Fpnu_def}
\end{equation}

There are several ways of solving this integral, as discussed in, e.g.~Ref.~\cite{Janik_2003}. Making the substitution $\mathbfss{Z} = \mathbfss{I} + i\mathbfss{T}$ in the definition of $P_{p,\nu}(\boldsymbol{\Omega} )$ gives
\begin{equation}
    P_{p,\nu}(\boldsymbol{\Omega} ) = i^{-\frac{p(p+1)}{2}} e^{-\frac{1}{2}\mathrm{Tr}(\boldsymbol{\Omega})} \int_{\mathrm{Re}(\mathbfss{Z})=\mathbfss{I}} \mathrm{d}\mathbfss{Z} \, e^{\frac{1}{2}\mathrm{Tr}(\boldsymbol{\Omega}\mathbfss{Z})} \lvert \mathbfss{Z} \rvert^{-\frac{\nu}{2}},
\end{equation}
where the integral is over a fixed real part and a symmetric imaginary part. The integral is an inverse Laplace transform, and can be derived from the result~\cite{10.2307/1969810}
\begin{equation}
    \frac{2^{\frac{1}{2}p(p-1)}}{(2\pi i)^{\frac{1}{2}p(p+1)}} \int_{\mathrm{Re}(Z) = X_0 > 0}\mathrm{d}\mathbfss{Z} \, e^{\mathrm{Tr}(\mathbfss{Z} \boldsymbol{\Lambda})} \lvert \mathbfss{Z} \rvert^{-a -\frac{1}{2}(p+1)} = \frac{\lvert \boldsymbol{\Lambda} \rvert^a}{\Gamma_p\left[a + \frac{1}{2}(p+1)\right]},
\end{equation}
where $\boldsymbol{\Lambda}>0$ and the integral is over symmetric imaginary parts and a fixed positive definite real part. We require $\boldsymbol{\Omega}> 0$, i.e.~$\nu \geq p$. Since $X_0 = I$ is positive definite, we get, after some rearranging,
\begin{equation}
    P_{p,\nu}(\boldsymbol{\Omega} ) = \frac{2^\frac{p(p-\nu+3)}{2}\pi^{\frac{p}{2}(p+1)}}{\Gamma_p\left(\frac{\nu}{2}\right)}  \left \vert \boldsymbol\Omega \right\vert^{\frac{\nu-p-1}{2}} e^{-\frac{1}{2}\mathrm{Tr}( \boldsymbol{\Omega})}.
    \label{eq:Fpnu}
\end{equation}

Substituting $P_{p,\nu}(\boldsymbol{\Omega} )$ back in to Equation~\eqref{eq:Wishart_char_pne1} gives
\begin{equation}
    p(\{\nu \hat{\mathbfss{C}}_l\}) =  \prod_{l=l_{\mathrm{min}}}^{l_{\mathrm{max}}} \frac{ \lvert \nu \hat{\mathbfss{C}}_l \rvert^{(\nu - p -1)/2}}{2^{\frac{p\nu}{2}}  \Gamma_p\left(\frac{\nu}{2}\right) \left \vert \mathbfss{C}_l \right \vert^{\frac{\nu}{2}}}e^{-\frac{\nu}{2}\mathrm{Tr}(\mathbfss{C}_l^{-1}\hat{\mathbfss{C}}_l)},
\end{equation}
i.e.~
\begin{align}
    p(\{ \hat{\mathbfss{C}}_l\}) &=  \prod_{l=l_{\mathrm{min}}}^{l_{\mathrm{max}}} \frac{ \lvert  \hat{\mathbfss{C}}_l \rvert^{(\nu - p -1)/2}}{2^{\frac{p\nu}{2}}  \Gamma_p\left(\frac{\nu}{2}\right) \left \vert \mathbfss{C}_l /\nu \right \vert^{\frac{\nu}{2}}}e^{-\frac{\nu}{2}\mathrm{Tr}(\mathbfss{C}_l^{-1}\hat{\mathbfss{C}}_l)} \nonumber \\
    &\equiv \prod_{l=l_{\mathrm{min}}}^{l_{\mathrm{max}}} W_p(\hat{\mathbfss{C}}_l; \mathbfss{C}_l/\nu, \nu),
\end{align}
where $W_p$ is the $p$-dimensional Wishart density.

\subsection{Convergence of the Wishart distribution to Gaussianity}
\label{appsubsec:Wish2inf}

The convergence of probability distributions is most easily studied at the level of the characteristic function. A single $\hat{\mathbfss{C}}_l$ has the characteristic function
\begin{equation}
    \phi_{\hat{\mathbfss{C}}_l}(\mathbfss{J}_l) = \left \vert \mathbfss{I} + \frac{2i\mathbfss{J}_l \mathbfss{C}_l}{\nu} \right \vert^{-\frac{\nu}{2}}.
\end{equation}
Consider the characteristic function of $\mathbfss{M}_l \equiv \mathbfss{C}^{-1/2}\hat{\mathbfss{C}}_l \mathbfss{C}_l^{-1/2}$. This is 
 \begin{equation}
    \phi_{\mathbfss{M}_l}(\mathbfss{J}_l) = \left \vert \mathbfss{I} + \frac{2i\mathbfss{J}_l }{\nu} \right \vert^{-\frac{\nu}{2}}.
\end{equation}  
We can rewrite this as
\begin{equation}
    \phi_{\mathbfss{M}_l}(\mathbfss{J}_l) = \left \vert \mathbfss{I} + \frac{2i \mathbfss{R} \mathbfss{J}_l  \mathbfss{R}^T}{\nu} \right \vert^{-\frac{\nu}{2}}.
\end{equation}
where $\mathbfss{R}$ is any orthogonal matrix. Choosing $\mathbfss{R}$ as the matrix that diagonalizes $\mathbfss{J}_l $, this gives
\begin{equation}
    \phi_{\mathbfss{M}_l}(\mathbfss{J}_l) = \prod_{i=1}^p \left [1 + \frac{2i \lambda_i}{\nu} \right ]^{-\frac{\nu}{2}}.
\end{equation}
where $\lambda_i$ are the eigenvalues of $\mathbfss{J}_l $.

Now fixing $\mathbfss{J}_l$, taking $\nu \to \infty$  and keeping terms up to $\mathcal{O}(\nu^{-1})$ gives
\begin{align}
    \lim_{\nu \to \infty} \phi_{\mathbfss{M}_l}(\mathbfss{J}_l) &= \exp\left(-i\sum_{i=1}^p \lambda_i - \sum_{i=1}^p \frac{\lambda_i^2}{\nu}\right) \nonumber \\
    &= \exp\left[ -i\mathrm{Tr}(\mathbfss{J}_l ) - \frac{1}{\nu} \mathrm{Tr}(\mathbfss{J}_l \mathbfss{J}_l) \right]
\end{align}
This implies that the characteristic function of our original variable is
\begin{equation}
    \lim_{\nu \to \infty} \phi_{\hat{\mathbfss{C}}_l}(\mathbfss{J}_l) = \exp\left[ -i\mathrm{Tr}(\mathbfss{J}_l \mathbfss{C}_l) - \frac{1}{\nu} \mathrm{Tr}(\mathbfss{J}_l \mathbfss{C}_l\mathbfss{J}_l\mathbfss{C}_l) \right].
\end{equation}
The inverse Fourier transform is a Gaussian integral, giving
\begin{equation}
    \lim_{\nu \to \infty} p(\hat{\mathbfss{C}}_l) = 2^{\frac{p(p-1)}{2}} \int \frac{\mathrm{d}\mathbfss{J}_l}{(2\pi)^{\frac{p(p+1)}{2}}} e^{i\mathrm{Tr}[\mathbfss{J}_l(\hat{\mathbfss{C}}_l - \mathbfss{C}_l)]} e^{-\frac{1}{\nu} \mathrm{Tr}(\mathbfss{J}_l \mathbfss{C}_l \mathbfss{J}_l \mathbfss{C}_l)}.
\end{equation}

To do this integral, it is advantageous to move to vectorized notation~\cite{Gupta}. For a general $p \times p$ matrix $Y$ define $\mathrm{vec}(Y)$ as the $p^2$-dimensional vector whose elements are $(Y_{1,1}, \dots, Y_{p,1}, Y_{1,2}, \dots, Y_{p,2}, \dots)$. For a symmetric $p \times p$ matrix $X$, define $\mathrm{vecp}(X)$ as the $\frac{1}{2}p(p+1)$-dimensional vector whose elements are $(X_{1,1}, X_{1,2}, \dots, X_{1,p}, X_{2,2}, X_{2,3}, \dots)$. Then some standard matrix identities give
\begin{align}
    \mathrm{Tr}(\mathbfss{X} \mathbfss{D} \mathbfss{X} \mathbfss{E}) &= \mathrm{vec}^T(\mathbfss{X})(\mathbfss{E} \otimes \mathbfss{D}) \mathrm{vec}(\mathbfss{X}) \nonumber \\
    &= \mathrm{vecp}^T(\mathbfss{X})\mathbfss{B}_p^+(\mathbfss{E} \otimes \mathbfss{D})\mathbfss{B}_p^{+ T} \mathrm{vecp}(\mathbfss{X}),
\end{align}
where $\otimes$ is the Kronecker product, and $\mathbfss{B}_p^+ = (\mathbfss{B}_p^T \mathbfss{B}_p)^{-1} \mathbfss{B}_p^T$ is the Moore-Penrose inverse of $\mathbfss{B}_p$, the $p^2 \times \frac12 p(p+1)$ transition matrix. This matrix satisfies $\mathrm{vec}(\mathbfss{X}) = \mathbfss{B}_p^{+ T} \mathrm{vecp}(\mathbfss{X})$ and $\mathrm{vec}(\mathbfss{X}) = \mathbfss{B}_p^T \mathrm{vecp}(\mathbfss{X})$, and has typical elements given by~\cite{Gupta}
\begin{equation}
(\mathbfss{B}_p)_{ij,gh} = \frac{1}{2}(\delta_{ig}\delta_{jh} + \delta_{ih}\delta_{jg}), \, i\leq p, \, j\leq p, \, g\leq h \leq p.
\end{equation}
Note that $\mathbfss{B}_p^+$ is a $\frac{1}{2}p(p+1) \times p^2$ matrix. We also have
\begin{align}
    \mathrm{Tr}(\mathbfss{X}\mathbfss{A}) &= \mathrm{vec}^T(\mathbfss{X}) \mathrm{vec}(\mathbfss{A}) \nonumber \\
    &= \mathrm{vecp}^T(\mathbfss{X})\mathbfss{B}_p^+ \mathbfss{B}_p^{+ T}\mathrm{vecp}(\mathbfss{A}).
\end{align}
Making these substitutions, defining $\Delta \hat{\mathbfss{C}}_l = \hat{\mathbfss{C}}_l - \mathbfss{C}_l$, and using that $\lvert \mathbfss{B}_p^T \mathbfss{B}_p \rvert = 2^{-\frac12 p(p-1)}$ gives
\begin{align}
    \lim_{\nu \to \infty} p(\hat{\mathbfss{C}}_l) &= 2^{\frac{p(p-1)}{2}} \int \frac{\mathrm{dvecp}(\mathbfss{J}_l)}{(2\pi)^{\frac{p(p+1)}{2}}} e^{i \mathrm{vecp}^T(\mathbfss{J}_l)\mathbfss{B}_p^+ \mathbfss{B}_p^{+ T}\mathrm{vecp}(\Delta \hat{\mathbfss{C}}_l) }  e^{-\frac{1}{\nu} \mathrm{vecp}^T(\mathbfss{J}_l)\mathbfss{B}_p^+(\mathbfss{C}_l \otimes \mathbfss{C}_l)\mathbfss{B}_p^{+ T} \mathrm{vecp}(\mathbfss{J}_l)} \nonumber \\
    &= \frac{\exp\left \{-\frac{1}{2} \mathrm{vecp}^T(\Delta \hat{\mathbfss{C}}_l) \left[\frac{2}{\nu}\mathbfss{B}_p^T (\mathbfss{C}_l \otimes \mathbfss{C}_l) \mathbfss{B}_p\right]^{-1} \mathrm{vecp}(\Delta \hat{\mathbfss{C}}_l)\right\}}{\sqrt{(2\pi)^{\frac{1}{2}p(p+1)} \left \vert \frac{2}{\nu} \mathbfss{B}_p^T (\mathbfss{C}_l \otimes \mathbfss{C}_l) \mathbfss{B}_p\right \vert}},
\end{align}
i.e.~$\mathrm{vecp}(\hat{\mathbfss{C}}_l)$ is $\frac12 p(p+1)$-dimensional Gaussian with mean $\mathrm{vecp}(\mathbfss{C}_l)$ and covariance matrix $\frac{2}{\nu} \mathbfss{B}_p^T (\mathbfss{C}_l \otimes \mathbfss{C}_l) \mathbfss{B}_p$. The matrix $\hat{\mathbfss{C}}_l$ is thus symmetric matrix normal. Note that the mean and covariance matrix do not change compared with their Wishart forms when the limit is taken.

We have shown that the characteristic function of the Wishart distribution tends pointwise to that of a symmetric matrix Gaussian in the limit $\nu \to \infty$. This is enough to establish that Wishart matrices converge in distribution to symmetric Gaussian matrices, by L\'{e}vy's continuity theorem.

\subsection{Signal with non-zero trispectrum}
\label{appsubsec:trispec}

We write the signal characteristic function as
\begin{equation}
     \phi_s(\{\mathbf{k}_{lm}\}) = \left(1 + \frac{1}{24} \kappa^{ijkm}_{\underline{lm}} k^i_{l_1 m_1} k^j_{l_2 m_2} k^k_{l_3 m_3}k^m_{l_4 m_4} \right) \phi_s^G(\{\mathbf{k}_{lm}\}).
\end{equation}
Substituting this into Equation~\eqref{eq:char_1} gives a Gaussian integral, as in the Gaussian signal case, but now with four factors of the wave number in the integrand. This integral can be done analytically, which gives the power spectrum characteristic function as 
\begin{equation}
    \phi_{\{\nu \hat{\mathbfss{C}}_l\}}(\{\mathbfss{J}_l\}) = \left(\prod_{l=l_{\mathrm{min}}}^{l_{\mathrm{max}}} \lvert \mathbfss{I} + 2i\mathbfss{J}_l\mathbfss{C}_l  \rvert^{-\frac{\nu}{2}}\right) \left[1 + \frac{1}{8}\sum_{l_1,l_2}(2l_1+1)(2l_2+1) \langle \Delta \hat{C}^{ij}_{l_1} \Delta \hat{C}^{km}_{l_2} \rangle_{\mathrm{NG}} M^{ij}_{l_1}M^{km}_{l_2}\right].
    \label{eq:char_with_tri_pne1}
\end{equation}

A few consequences of the corrected distribution are immediately evident from Equation~\eqref{eq:char_with_tri_pne1}. Firstly, the leading-order correction in the limit $J_l \to 0$ is quadratic in $J_l$, which tells us that both the normalization and the mean of $p(\hat{C}_l)$ are unaffected by our correction (the latter is true by construction), but the covariance does receive a correction, given by $\langle \Delta \hat{C}^{ij}_{l_1} \Delta \hat{C}^{km}_{l_2} \rangle_{\mathrm{NG}}$. This is as expected, agreeing with the non-perturbative expression Equation~\eqref{eq:NGcov}.

Inverse Fourier-transforming Equation~\eqref{eq:char_with_tri_pne1} gives the correction to the probability density as
\begin{equation}
    \Delta p(\{\nu \hat{\mathbfss{C}}_l\}) = \frac{1}{8}2^{\frac{\lambda p(p-1)}{2}} \int \frac{\mathrm{d}\{\mathbfss{J}_l\}}{(2\pi)^{n\lambda}} e^{i\sum_l \mathrm{Tr}(\nu\mathbfss{J}_l \hat{\mathbfss{C}}_l)} \left(\prod_{l=l_{\mathrm{min}}}^{l_{\mathrm{max}}} \lvert \mathbfss{I} + 2i\mathbfss{J}_l\mathbfss{C}_l  \rvert^{-\frac{\nu}{2}}\right) \sum_{l_1,l_2}\nu_1\nu_2 \langle \Delta \hat{C}^{ij}_{l_1} \Delta \hat{\mathrm{C}}^{km}_{l_2} \rangle_{\mathrm{NG}}(\mathbfss{M}_{l_1})^{ij}(\mathbfss{M}_{l_2})^{km},
    \label{eq:Dp_1_pne1}
\end{equation}
with $\mathbfss{M}_l = [\mathbfss{C}_l + (2i\mathbfss{J}_l)^{-1}]^{-1}$ and $\lambda \equiv l_{\mathrm{max}} - l_{\mathrm{min}} + 1$. The quantity $\langle \Delta \hat{C}^{ij}_{l_1} \Delta \hat{C}^{km}_{l_2} \rangle_{\mathrm{NG}}$ is the non-Gaussian contribution to the covariance matrix. In the $p=1$ case we call this $T_{l_1 l_2}$. The crucial effect of non-Gaussianity here is to couple $\hat{C}_l$ with different $l$, a result of non-zero off-diagonal terms in the non-Gaussian covariance matrix; the probability density no longer factorizes in $l$. 

The integrals in Equation~\eqref{eq:Dp_1_pne1} are of the Wishart type when $l_1 \neq l_2$, and the only new term is that with $l_1 = l_2$. After some rearranging and relabelling, one can show that it is necessary to compute the integral
\begin{equation}
     H^{ab,cd}_{p,\nu}(\boldsymbol{\Omega}) \equiv \int \mathrm{d}\mathbfss{T} \, e^{\frac{i}{2}\mathrm{Tr}(\boldsymbol{\Omega}\mathbfss{T})} \, \lvert \mathbfss{I} + i\mathbfss{T} \rvert^{-\frac{\nu}{2}} \left[(\mathbfss{I} + i\mathbfss{T})^{-1} \right]_{ab}\left[ (\mathbfss{I} + i\mathbfss{T})^{-1} \right]_{cd}.
     \label{eq:Hpnu_def}
\end{equation}
We'll evaluate this using derivatives, but first we need to establish a few rules for taking derivatives with respect to symmetric matrices. These can be found in Ref.~\cite{Petersen2008}.

For a symmetric matrix $\mathbfss{X}$, we have
\begin{align}
    \frac{\partial \mathbfss{X}}{\partial X_{ij}} &= \mathbfss{J}^{ij} + \mathbfss{J}^{ji} - \mathbfss{J}^{ij} \mathbfss{J}^{ij} \\
    \frac{\partial \mathrm{Tr}(\mathbfss{A}\mathbfss{X})}{\partial \mathbfss{X}} &= \mathbfss{A} + \mathbfss{A}^T - (\mathbfss{A} \circ \mathbfss{I}) \\
    \frac{\partial \mathrm{det}\mathbfss{X}}{\partial \mathbfss{X}} &= (\mathrm{det}\mathbfss{X}) [2\mathbfss{X}^{-1} - (\mathbfss{X}^{-1} \circ \mathbfss{I})],
\end{align}
where $\circ$ denotes the Hadamard (element-wise) product, and $\mathbfss{J}^{ij}$ is the matrix whose $km$ element is $\delta_{ik}\delta_{jm}$, i.e.~the matrix whose only non-zero values are in cell $ij$.

We can therefore write the following identity holding for symmetric matrices $\mathbfss{T}$:
\begin{equation}
    \lvert \mathbfss{I} + i\mathbfss{T} \rvert^{-\frac{\nu}{2}} (\mathbfss{I} + i\mathbfss{T})^{-1}_{ij} = \frac{i}{\nu} \frac{\partial}{\partial T_{ij}} \lvert \mathbfss{I} + i\mathbfss{T} \rvert^{-\frac{\nu}{2}} + \frac{1}{2} (\mathbfss{I} + i\mathbfss{T})^{-1}_{ij} \delta_{ij},
\end{equation}
with \emph{no summation over repeated indices}. Using this identity one can establish the identity
\begin{align}
    \lvert \mathbfss{I} + i\mathbfss{T} \rvert^{-\frac{\nu}{2}} (\mathbfss{I} + i\mathbfss{T})^{-1}_{ij}(\mathbfss{I} + i\mathbfss{T})^{-1}_{km} &= \frac{i}{\nu} \frac{\partial}{\partial T_{ij}} \left[ \lvert \mathbfss{I} + i\mathbfss{T} \rvert^{-\frac{\nu}{2}} (\mathbfss{I} + i\mathbfss{T})^{-1}_{km}\right] + \frac{1}{2} \lvert \mathbfss{I} + i\mathbfss{T} \rvert^{-\frac{\nu}{2}} (\mathbfss{I} + i\mathbfss{T})^{-1}_{ij} \delta_{ij} (\mathbfss{I} + i\mathbfss{T})^{-1}_{km} \nonumber \\
    & - \frac{1}{\nu} \lvert \mathbfss{I} + i\mathbfss{T} \rvert^{-\frac{\nu}{2}} \left[2 (\mathbfss{I} + i\mathbfss{T})^{-1}_{i\left(k\right.} (\mathbfss{I} + i\mathbfss{T})^{-1}_{m\left. \right)j} - \delta_{ij} (\mathbfss{I} + i\mathbfss{T})^{-1}_{kj} (\mathbfss{I} + i\mathbfss{T})^{-1}_{mi}\right],
\end{align}
where again there is no summation over repeated indices, and parentheses denote symmetrization over enclosed indices. Substituting this in to Equation~\eqref{eq:Hpnu_def}, integrating by parts and noting that the boundary term vanishes (which it must do for the Fourier transforms to converge) gives
\begin{equation}
    H^{ijkm}_{p,\nu}(\boldsymbol{\Omega}) = \frac{1}{2\nu} (2\boldsymbol{\Omega} - \boldsymbol{\Omega} \circ \mathbfss{I})_{ij}P_{p,\nu}(\boldsymbol{\Omega})\frac{\Omega_{km}}{\nu} + \frac{1}{2}\delta_{ij}H^{ijkm}_{p,\nu}(\boldsymbol{\Omega})  - \frac{2}{\nu} H^{i(km)j}_{p,\nu}(\boldsymbol{\Omega}) + \frac{1}{\nu}\delta_{ij} H^{kjmi}_{p,\nu}(\boldsymbol{\Omega}).
\end{equation}
Permuting indices we can get three simultaneous equations for $H^{ijkm}_{p,\nu}$. The solution is
\begin{equation}
    H^{ijkm}_{p,\nu}(\boldsymbol{\Omega}) = \frac{(\nu+1)}{\nu(\nu+2)(\nu-1)} \left[\Omega_{ij}\Omega_{km} - \frac{2}{(\nu+1)} \Omega_{i\left(k \right.}\Omega_{\left. m\right)j} \right] P_{p,\nu}(\boldsymbol{\Omega}).
    \label{eq:Hpnu_sol}
\end{equation}
Putting everything together gives the corrected density for the $\hat{\mathbfss{C}}_l$ as
\begin{align}
    p(\{\hat{\mathbfss{C}}_l \}) &= \left[\prod_{l=l_{\mathrm{min}}}^{l_{\mathrm{max}}} W_p(\hat{\mathbfss{C}}_l; \mathbfss{C}_l/\nu, \nu)\right] \left\{1 + \frac{1}{8}\sum_{l_1, l_2} \nu_1 \nu_2 \langle \Delta \hat{C}^{ab}_{l_1} \Delta \hat{C}^{cd}_{l_2} \rangle_{\mathrm{NG}} C^{-1}_{l_1,ai} C^{-1}_{l_1,bj} C^{-1}_{l_2,ck} C^{-1}_{l_2,dm} \vphantom{\frac12} \right. \nonumber \\
    & \times \left. \left[ \Delta \hat{C}_{l_1,ij}\Delta \hat{C}_{l_2,km} + \frac{2\delta_{l_1 l_2}}{(\nu+2)(\nu-1)} \left(\hat{C}_{l_1,ij}\hat{C}_{l_1,km} - \nu_1 \hat{C}_{l_1,i\left(k \right.}\hat{C}_{l_1,\left. m\right)j}\right)\right] \right\},
    \label{eq:finalpCl_tri_pne1}
\end{align}
with implicit summation over repeated indices. One can verify that the distribution Equation~\eqref{eq:finalpCl_tri_pne1} is correctly normalised and has a mean given by $\langle \hat{\mathbfss{C}}_l \rangle = \mathbfss{C}_l$, i.e.~unchanged from the Wishart case. 

We can write the correction in a more compact vectorial notation as follows. We define the Gaussian covariance matrix 
\begin{equation}
    \boldsymbol{\Sigma}_l \equiv \langle \mathrm{vecp}(\Delta \hat{\mathbfss{C}}_l) \mathrm{vecp}^T(\Delta \hat{\mathbfss{C}}_l) \rangle_G = \frac{2}{\nu} \mathbfss{B}_p^T (\mathbfss{C}_l \otimes \mathbfss{C}_l) \mathbfss{B}_p,
\end{equation}
which implies that $\boldsymbol{\Sigma}^{-1}_l = \frac{\nu}{2} \mathbfss{B}_p^+ (\mathbfss{C}_l^{-1} \otimes \mathbfss{C}_l^{-1}) \mathbfss{B}_l^{+ T}$. After some manipulations, the term quadratic in $\Delta \hat{C}_l$ in Equation~\eqref{eq:finalpCl_tri} can be written as
\begin{equation}
    \frac{1}{2} \sum_{l_1, l_2} \mathrm{vecp}^T(\Delta \hat{\mathbfss{C}}_{l_1}) \boldsymbol{\Sigma}_{l_1}^{-1} \langle \mathrm{vecp}(\Delta \hat{\mathbfss{C}}_{l_1})\mathrm{vecp}(\Delta \hat{\mathbfss{C}}_{l_2})\rangle_{\mathrm{NG}} \boldsymbol{\Sigma}_{l_2}^{-1} \mathrm{vecp}^T(\Delta \hat{\mathbfss{C}}_{l_2}),
\end{equation}
which is the next-to-leading-order term in the expansion of 
\begin{equation}
    -\frac{1}{2} \sum_{l_1, l_2} \mathrm{vecp}^T(\Delta \hat{\mathbfss{C}}_{l_1}) \left[ \boldsymbol{\Sigma}_{l_1}\delta_{l_1l_2} + \langle \mathrm{vecp}(\Delta \hat{\mathbfss{C}}_{l_1})\mathrm{vecp}^T(\Delta \hat{\mathbfss{C}}_{l_2})\rangle_{\mathrm{NG}} \right]^{-1} \mathrm{vecp}^T(\Delta \hat{\mathbfss{C}}_{l_2}),
\end{equation}
i.e.~an expansion of a chi-squared having the inverse of the \emph{total} covariance matrix. Once the Gaussian part of this expansion is `available' when the $\nu \to \infty$ limit is taken, we will be left a Gaussian possessing the correct inverse covariance matrix in its exponent.

The first of the two terms quadratic in $\hat{C}_l$ in Equation~\eqref{eq:finalpCl_tri_pne1} has the same form as the term above, and can be written as a quadratic in $\mathrm{vecp}(\hat{\mathbfss{C}}_l)$. Note that this term is suppressed with respect to the second term quadratic in $\hat{C}_l$ by a factor of $\nu$. This second term can be written as, with $\mathbfss{A}_l \equiv \mathbfss{C}_l^{-1} \hat{\mathbfss{C}}_l \mathbfss{C}_l^{-1}$
\begin{align}
    &-\frac{1}{4}\sum_{l_1} \frac{\nu_1^3}{(\nu_1+2)(\nu_1-1)}\mathrm{Tr}\left[(\mathbfss{A}_{l_1} \otimes \mathbfss{A}_{l_1}) \langle \mathrm{vec}(\Delta \hat{\mathbfss{C}}_{l_1}) \mathrm{vec}^T(\Delta \hat{\mathbfss{C}}_{l_1})\rangle_{\mathrm{NG}} \right] \nonumber \\
    &= -\frac{1}{4}\sum_{l_1} \frac{\nu_1^3}{(\nu_1+2)(\nu_1-1)}\mathrm{Tr}\left[\mathbfss{B}_p^+(\mathbfss{A}_{l_1} \otimes \mathbfss{A}_{l_1})\mathbfss{B}_p^{+T} \langle \mathrm{vecp}(\Delta \hat{\mathbfss{C}}_{l_1}) \mathrm{vecp}^T(\Delta \hat{\mathbfss{C}}_{l_1})\rangle_{\mathrm{NG}} \right].
\end{align}
This looks like the next-to-leading-order term in an expansion of the determinant of the total covariance, with the exception that it depends on the \emph{measured} power. In the limit $\nu_1 \to \infty$ we can write $\hat{\mathbfss{C}}_l = \mathbfss{C}_l + \mathcal{O}(\nu^{-1})$, so we set $\hat{\mathbfss{C}}_l = \mathbfss{C}_l$ in this term and hence $\mathbfss{A}_l = \mathbfss{C}_l^{-1}$. In that limit, this term dominates over the first term quadratic in $\hat{C}_l$ and is precisely the expansion of the determinant of the \emph{total} covariance, i.e.~the prefactor in the asymptotic Gaussian density.

\subsection{Signal with non-zero bispectrum}
\label{appsubsec:bispec}

We now consider a signal having non-zero bispectrum. The signal characteristic function reads
\begin{equation}
    \phi_s(\{\mathbf{k}_{lm}\}) = \left(1  - \frac{1}{72} \kappa^{ijk}_{\underline{lm}} \kappa^{lmn}_{\underline{l'm'}} k^i_{l_1 m_1} k^j_{l_2 m_2} k^k_{l_3 m_3}k^l_{l_1' m_1'} k^m_{l_2' m_2'} k^n_{l_3' m_3'} \right)\phi_s^G(\{\mathbf{k}_{lm}\}).
\end{equation}
Following the same steps as the trispectrum case, the correction to the characteristic function is 
\begin{align}
    \Delta \phi_{\{\nu \hat{\mathbfss{C}}_l\}}(\{\mathbfss{J}_l\}) =  &\left(\prod_{l=l_{\mathrm{min}}}^{l_{\mathrm{max}}} \lvert \mathbfss{I} + 2i\mathbfss{J}_l\mathbfss{C}_l  \rvert^{-\frac{\nu}{2}}\right) \nonumber \\
    &\times \left[1 -\frac{1}{12}\sum_{l_1,l_2, l_3}\nu_1 \nu_2 \nu_3 \tilde{B}^{ijk,lmn}_{l_1l_2l_3} \left(\mathbfss{C}_{l_1} -\frac{i}{2} \mathbfss{J}_{l_1}^{-1}\right)^{-1}_{il}\left(\mathbfss{C}_{l_2} - \frac{i}{2}\mathbfss{J}_{l_2}^{-1}\right)^{-1}_{jm}\left(\mathbfss{C}_{l_3} - \frac{i}{2}\mathbfss{J}_{l_3}^{-1}\right)^{-1}_{kn}\right].
    \label{eq:char_with_bi_pne1}
\end{align}
In the limit that $J_l \to 0$ this is a cubic correction to the Wishart characteristic function, telling us that the effect of $\tilde{B}$ is to modify the three-point function or skewness of $\hat{C}_l$, as expected. Just as in the trispectrum case, the correction is not a polynomial in $J_l$ and hence not an Edgeworth expansion, even though it is perturbatively `close' to the zero-order Wishart distribution\footnote{Note that to get the characteristic function of $\hat{C}_l$ rather than $\nu \hat{C}_l$ one has to replace $J_l$ with $J_l/\nu$.}.

Inverse-Fourier transforming Equation~\eqref{eq:char_with_bi_pne1} gives the correction to the density as
\begin{equation}
    \Delta p(\{\nu \hat{\mathbfss{C}}_l\}) = -\frac{1}{12}2^{\frac{\lambda p(p-1)}{2}} \int \frac{\mathrm{d}\{\mathbfss{J}_l\}}{(2\pi)^{n\lambda}} e^{i\sum_l \mathrm{Tr}(\nu\mathbfss{J}_l \hat{\mathbfss{C}}_l)} \left(\prod_{l=l_{\mathrm{min}}}^{l_{\mathrm{max}}} \lvert \mathbfss{I} + 2i\mathbfss{J}_l\mathbfss{C}_l  \rvert^{-\frac{\nu}{2}}\right) \sum_{l_1,l_2,l_3}\nu_1\nu_2\nu_3 \tilde{B}^{ijk,lmn}_{l_1l_2l_3}(\mathbfss{M}_{l_1})^{il}(\mathbfss{M}_{l_2})^{jm}(\mathbfss{M}_{l_3})^{kn},
    \label{eq:Dp_bi_pne1}
\end{equation}
with $\mathbfss{M}_l = [\mathbfss{C}_l + (2i\mathbfss{J}_l)^{-1}]^{-1}$.

Once again the only new term is the case $l_1 = l_2 = l_3$. The new integral we need is
\begin{equation}
     K^{ab,cd, ef}_{p,\nu}(\boldsymbol{\Omega}) \equiv  \int \mathrm{d}\mathbfss{T} \, e^{\frac{i}{2}\mathrm{Tr}(\boldsymbol{\Omega}\mathbfss{T})} \, \lvert \mathbfss{I} + i\mathbfss{T} \rvert^{-\frac{\nu}{2}} \left[(\mathbfss{I} + i\mathbfss{T})^{-1} \right]_{ab}\left[ (\mathbfss{I} + i\mathbfss{T})^{-1} \right]_{cd} \left[ (\mathbfss{I} + i\mathbfss{T})^{-1} \right]_{ef}.
     \label{eq:Kpnu_def}
\end{equation}
Our strategy will again be to take derivatives. Letting $\mathbfss{A} \equiv (\mathbfss{I} + i\mathbfss{T})^{-1}$, we have the identity
\begin{align}
    \lvert \mathbfss{A} \rvert^{\frac{\nu}{2}} A_{ab}A_{ij}A_{km} &= 
     \frac{i}{\nu}\frac{\partial}{\partial T_{ij}} \left( \lvert \mathbfss{A}\rvert^{\frac{\nu}{2}} A_{km}A_{ab}\right)
    + \frac{1}{2} \delta_{ij} \lvert \mathbfss{A} \rvert^{\frac{\nu}{2}} A_{ab}A_{ij}A_{km} \nonumber \\
    & - \frac{2}{\nu}\lvert \mathbfss{A}\rvert^{\frac{\nu}{2}} A_{ab}A_{k\left(i \right.}A_{\left. j\right) m} + \frac{1}{\nu}\delta_{ij} \lvert \mathbfss{A} \rvert^{\frac{\nu}{2}} A_{ab}A_{ki}A_{mj} - \frac{2}{\nu}\lvert \mathbfss{A}\rvert^{\frac{\nu}{2}} A_{km}A_{a\left(i \right.}A_{\left. j\right) b}  +  \frac{1}{\nu}\delta_{ij} \lvert \mathbfss{A} \rvert^{\frac{\nu}{2}} A_{km}A_{ai}A_{jb},
\end{align}
with no summation over repeated indices.

Integrating by parts, substituting Equation~\eqref{eq:Hpnu_sol}, and defining $K^{ab,ij,km}_{p,\nu} \equiv M^{ab,ij,km}P_{p,\nu}$ (where $P_{p,\nu}$ is defined in Equation~\eqref{eq:Fpnu}) gives
\begin{align}
    &M^{ab,ij,km} - \frac{1}{2}\delta_{ij}M^{ab,ij,km} + \frac{2}{\nu}M^{ab,k(i,j)m} - \frac{1}{\nu}\delta_{ij} M^{ab,ki,mj} + \frac{2}{\nu}M^{km,a(i,j)b} 
     - \frac{1}{\nu}\delta_{ij}M^{km,ai,jb} \nonumber \\
     &= \frac{(\nu+1)}{\nu^2(\nu+2)(\nu-1)}\left(\Omega_{ij} - \frac{1}{2}\Omega_{ij}\delta_{ij}\right) \left(\Omega_{km}\Omega_{ab} - \frac{2}{\nu+1}\Omega_{k\left(a\right.}\Omega_{\left. b\right)m}\right).
\end{align}
Multiplying by $\delta_{ij}$ and subtracting the result gives
\begin{equation}
    M^{ab,ij,km}  + \frac{2}{\nu}M^{ab,k(i,j)m} + \frac{2}{\nu}M^{km,a(i,j)b} = \frac{(\nu+1)}{\nu^2(\nu+2)(\nu-1)} \left(\Omega_{ij}\Omega_{km}\Omega_{ab} - \frac{2}{\nu+1}\Omega_{ij}\Omega_{k\left(a\right.}\Omega_{\left. b\right)m}\right).
     \label{eq:Mabijkm_pne1}
\end{equation}
We can write two further equations like Equation~\eqref{eq:Mabijkm_pne1} by taking derivatives with respect to $T_{km}$ and $T_{ab}$. Adding all three equations gives an expression manifestly symmetric in its indices
\begin{align}
    &3M^{ab,ij,km} + \frac{2}{\nu}\left(M^{ab,ki,jm} + M^{ab,kj,im} + M^{km,aj,ib} + M^{km,ai,jb} + M^{ij,ak,mb} +M^{ij,am,kb} \right) \nonumber \\
    &= \frac{(\nu+1)}{\nu^2(\nu+2)(\nu-1)}\left[3\Omega_{ij}\Omega_{km}\Omega_{ab} - \frac{\left(\Omega_{ij}\Omega_{ka}\Omega_{bm} + \Omega_{ij}\Omega_{kb}\Omega_{am} + \Omega_{km}\Omega_{ia}\Omega_{bj} + \Omega_{km}\Omega_{ib}\Omega_{aj} + \Omega_{ab}\Omega_{ki}\Omega_{jm} + \Omega_{ab}\Omega_{kj}\Omega_{im} \right)}{1+\nu} \right].
    \label{eq:M_sym_pne1}
\end{align}
From the $p=1$ case we know that the solution must be cubic in the elements of $\boldsymbol{\Omega}$. There are three kinds of term that we can use to build $M^{ab, ij, km}$ dictated by the symmetries in play, given by
\begin{align}
    M^{ab,ij,km} = &X \Omega_{ij}\Omega_{km}\Omega_{ab} \nonumber \\
    + 2&Y[\Omega_{ij}\Omega_{k\left(a\right.}\Omega_{\left.b\right)m} + \Omega_{km}\Omega_{i\left(a\right.}\Omega_{\left.b\right)j} + \Omega_{ab}\Omega_{k\left(i\right.}\Omega_{\left.j\right)m}] \nonumber \\
     +& 2Z[\Omega_{jm}\Omega_{k\left(a\right.}\Omega_{\left.b\right)i} + \Omega_{ik}\Omega_{j\left(a\right.}\Omega_{\left.b\right)m} + \Omega_{jk}\Omega_{i\left(a\right.}\Omega_{\left.b\right)m} + \Omega_{im}\Omega_{k\left(a\right.}\Omega_{\left.b\right)j}],
\end{align}
for unknown coefficients $X, Y, Z$. The three terms here are closed under the symmetries of $M^{ab,ij,km}$. Substituting in to Equation~\eqref{eq:M_sym_pne1} and identifying coefficients of these terms gives the following equations for the unknown coefficients
\begin{align}
    3X + \frac{12}{\nu}Y &= \frac{3(\nu+1)}{\nu^2(\nu+2)(\nu-1)} \\
    6Y + \frac{2}{\nu}(2X + 2Y + 8Z) &= -\frac{2}{\nu^2(\nu+2)(\nu-1)} \\
    6Z + \frac{2}{\nu}(6Y + 6Z) &= 0.
\end{align}
The solution is
\begin{align}
    X &=  \frac{\nu^2 + 3\nu - 2}{\nu(\nu-1)(\nu-2)(\nu+2)(\nu+4)} \\
    Y &= \frac{-1}{\nu(\nu-1)(\nu-2)(\nu+4)} \\
    Z &= \frac{2}{\nu(\nu-1)(\nu-2)(\nu+2)(\nu+4)}.
\end{align}

The integral Equation~\eqref{eq:Kpnu_def} is thus
\begin{align}
    &K_{p,nu}^{ab,cd,ef}(\boldsymbol{\Omega}) = \frac{P_{p,\nu}(\boldsymbol{\Omega})}{\nu(\nu+4)(\nu-2)(\nu-1)} \left\{\frac{\nu^2 + 3\nu -2}{\nu+2} \Omega_{ab}\Omega_{cd}\Omega_{ef} -  2[\Omega_{cd}\Omega_{e\left(a\right.}\Omega_{\left.b\right)f} + \Omega_{ef}\Omega_{c\left(a\right.}\Omega_{\left.b\right)d} + \Omega_{ab}\Omega_{e\left(c\right.}\Omega_{\left.d\right)f}] \right. \nonumber \\
    & \left. + \frac{4}{\nu+2}[\Omega_{df}\Omega_{e\left(a\right.}\Omega_{\left.b\right)c} + \Omega_{ce}\Omega_{d\left(a\right.}\Omega_{\left.b\right)f} + \Omega_{de}\Omega_{c\left(a\right.}\Omega_{\left.b\right)f} + \Omega_{cf}\Omega_{e\left(a\right.}\Omega_{\left.b\right)d}] \right \},
\end{align}

Putting everything together gives the correction to the density as
\begin{align}
    &p(\{\hat{\mathbfss{C}}_l \}) = \left[\prod_{l=l_{\mathrm{min}}}^{l_{\mathrm{max}}} W_p(\hat{\mathbfss{C}}_l; \mathbfss{C}_l/\nu, \nu)\right] \left\{1 + \frac{1}{12}\sum_{l_1, l_2,l_3} \nu_1 \nu_2 \nu_3 \tilde{B}_{l_1l_2l_3}^{abc,def} C^{-1}_{l_1,ai} C^{-1}_{l_2,bj} C^{-1}_{l_3,ck} C^{-1}_{l_1,dl} C^{-1}_{l_2,em} C^{-1}_{l_3,fn} \vphantom{\frac12} \right. \nonumber \\
    & \times \left[ \vphantom{\frac12} \Delta \hat{C}_{l_1,il} \Delta \hat{C}_{l_2,jm} \Delta \hat{C}_{l_3,kn} + [3]\frac{2\delta_{l_1 l_2}}{(\nu_1+2)(\nu_1-1)} \hat{C}_{l_1,il}\hat{C}_{l_1,jm} \Delta \hat{C}_{l_3,kn} -[3] \frac{2\delta_{l_1 l_2}\nu_1}{(\nu_1+2)(\nu_1-1)} \hat{C}_{l_1,i\left(j\right.}\hat{C}_{l_1,\left.m \right)l}\Delta \hat{C}_{l_3,kn} \right. \nonumber \\
    &+ \frac{32\delta_{l_1 l_2}\delta_{l_2 l_3}}{(\nu_1+2)(\nu_1-1)(\nu_1+4)(\nu_1-2)}\hat{C}_{l_1,il}\hat{C}_{l_1,jm}\hat{C}_{l_1,kn} - [3]\frac{16\nu_1\delta_{l_1 l_2}\delta_{l_2 l_3}}{(\nu_1+2)(\nu_1-1)(\nu_1+4)(\nu_1-2)}\hat{C}_{l_1,jm}\hat{C}_{l_1,k\left( i\right.}\hat{C}_{l_1,\left. l \right) n} \nonumber \\
    & \left. \left.+ [4]\frac{4\nu_1^2\delta_{l_1 l_2}\delta_{l_2 l_3}}{(\nu_1+2)(\nu_1-1)(\nu_1+4)(\nu_1-2)}\hat{C}_{l_1,mn}\hat{C}_{l_1,k\left( i\right.}\hat{C}_{l_1,\left. l\right)j}\right] \vphantom{\frac12}\right\},
    \label{eq:finalpCl_bisq_pne1}
\end{align}
with implicit summation over repeated indices. We have verified that this distribution is correctly normalised. Computing the corrections to the mean, covariance, and three-point function is tedious so we have only checked for the $p=1$ case, confirming that these three cumulants are recovered correctly.

Equation~\eqref{eq:finalpCl_tri_pne1} and Equation~\eqref{eq:finalpCl_bisq_pne1} give the leading order correction to the Wishart distribution from signal non-Gaussian. The expressions are clearly quite cumbersome when $p \neq 1$, but may be simplified using the symmetries of the bin indices. It may also be possible to write these expressions in a more covariant way, although we were not able to find a more compact expression.

\section{Alternative binning choices}
\label{app:dlog0p2}

The default results of Section~\ref{subsec:results} are made with evenly space logarithmically binned power spectra with $\Delta \log l = 0.1$ and a weight function $W_l = 2l+1$. The mathematical details of the binning are described in Section~\ref{subsec:binning}.

The main use of binning for our purposes is to allow us to easily measure the covariance matrix of our lognormal maps from a finite number of simulations. Binning reduces the Monte Carlo noise in this measurement and increases the relative strength of the non-Gaussian part. 

\begin{figure}
\centering
\includegraphics[width=\columnwidth]{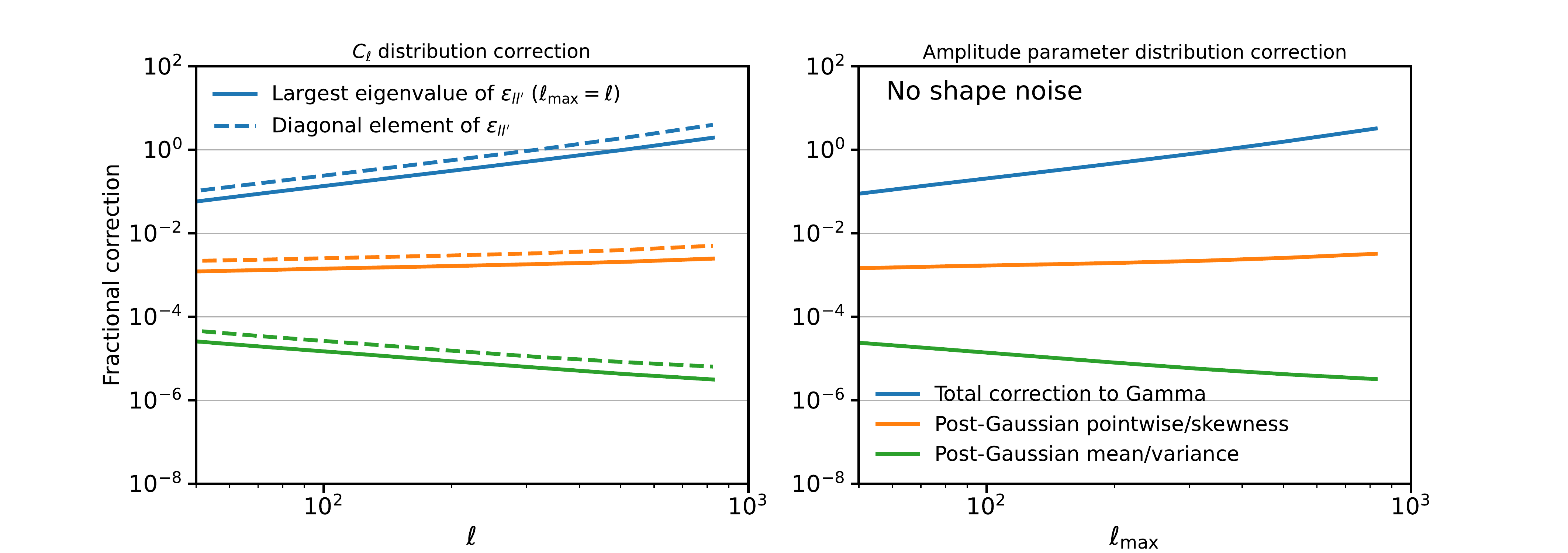}
\caption{The leading-order fractional corrections to the Gamma posterior with bandpowers binned with $\Delta \log l = 0.2$. Curves have the same meaning as in Figure~\ref{fig:epsmat}.}
\label{fig:epsmat_0p2}
\end{figure}

As discussed in Section~\ref{subsec:binning}, the correction to posteriors from signal non-Gaussianity in expected to depend only mildly on the bin width, as we verify numerically in Figure~\ref{fig:epsmat_0p2} where we plot the corrections to $C_l$ and amplitude posteriors for logarithmic bins with $\Delta \log l = 0.2$. This plot should be compared with Figure~\ref{fig:epsmat}. The impact of these broader bins is very mild. The broader bins roughly preserve the total mode count when $W_l = 2l+1$, so the effect on the posterior is naturally expected to be small.
%

It is beyond the scope of this paper to provide an exhaustive investigation of the different binning choices for the power spectrum, but a few obvious schemes deserve discussion. Instead of logarithmic bins we can also choose linearly-spaced bins. This choice compresses many of the large-scale modes into a single bin and hence does not allow the low-$l$ behaviour of the likelihood to be clearly seen, so we chose not to adopt this choice in our baseline results. Another alternative is to use a weight function $W_l = l(l+1)$, which is a common choice for power spectrum bandpowers. This upweights the smallest scales in each bin and consequently causes our likelihood approximation to break down at smaller $l_b$ (the bin barycentre) for logarithmic bins with $\Delta l = 0.1$. An advantage of our formalism is that the likelihood correction diagnostic $\langle \varepsilon \rangle$ can be computed for any binning choice, something which we advocate when analysing power spectra binned with this weight function.

We close by noting that any binning choice may be adopted so long as the bins are sufficiently small that the power spectrum can be approximated as flat over the extent of the bin. This assumption can be avoided entirely by working with $\hat{C}_l/C_l$ rather than $C_l$, as discussed in Section~\ref{subsec:binning}.

\section{Dependence on source redshift}
\label{app:zs6}

The baseline results in this paper have assumed a broad source redshift bin following a Euclid-like distribution, peaking at $z_s \approx 1$. Lensing surveys typically place galaxies in tomographic redshift bins - the fiducial choice of Ref.~\cite{2020A&A...642A.191E} uses 10 redshift bins, which is representative of upcoming Stage-IV surveys. Assuming equipopulated bins, the redshift distribution of the lowest bin is approximately Gaussian (after accounting for photometric redshift errors) with a peak around $z_s = 0.325$ and width $\sigma_z = 0.1$. Non-Gaussianity in the shear field is relatively stronger at a fixed angular scale at lower redshift due to that scale subtending smaller, more non-linear spatial scales and the evolution of structure formation towards lower redshift. This is offset to an extent by the overall signal strength being suppressed by the lensing kernel compared with a bin at higher source redshift. This also makes the relative effect of any Gaussianizing shape noise stronger.

To test our likelihood approximation in what is likely the most extreme case for a Euclid-like survey, we generate an ensemble of lognormal fields for this low-redshift source bin. We scale the `shift' parameter quantifying the strength of the lognormality according to the redshift-scaling formula proposed by Ref.~\cite{2016MNRAS.459.3693X}, confirming this provides a reasonable match to the covariance matrix of the ray-traced N-body simulations of Ref.~\cite{2017ApJ...850...24T}.

\begin{figure}
\centering
\includegraphics[width=\columnwidth]{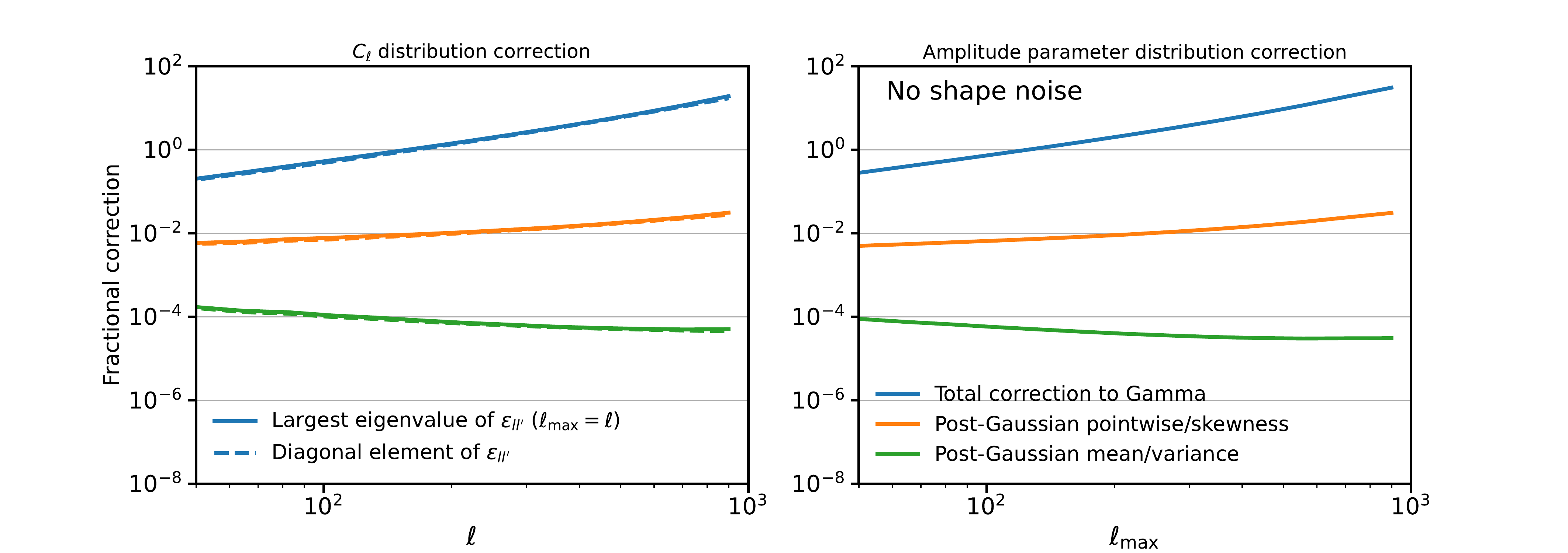}
\caption{Leading-order fractional corrections to the Gamma posterior for a Gaussian source redshift distribution centered at $z_s = 0.325$ with width $\sigma_z = 0.1$. Power spectra are assumed to be binned with $\Delta \log l = 0.1$. Curves have the same meaning as in Figure~\ref{fig:epsmat}.}
\label{fig:epsmat_zs6}
\end{figure}

Figure~\ref{fig:epsmat_zs6} shows the leading-order corrections to the posterior and the convergence statistics for this redshift bin, which should be compared with Figure~\ref{fig:epsmat}. The stronger non-Gaussianity is clear to see, with the blue curves quantifying the correction to the Fisher matrix on an individual $C_l$ (left panel) or an amplitude parameter (right panel) passing unity around $l \approx 100$. Our assumption of perturbative non-Gaussianity in the signal is valid only up to around $l \approx 30$. At this angular scale the Gamma distribution is reasonably well approximated by a Gaussian, and passing the full non-Gaussian covariance matrix to this Gaussian absorbs most of the correction to the posterior that our approximation implies. Residual post-Gaussian effects are at most 1\%, and extrapolating out to $l = 1000$ we see that corrections are likely to be no more than 10\%. It therefore appears that even for this low source redshift bin the scale of non-Gaussianity ($l \approx 30$) and the scale where Gamma-type corrections are necessary ($l \approx 10$) are separated enough that the likelihood can be approximated as transitioning from a Gamma to a Gaussian with the correct covariance matrix.

\begin{figure}
\centering
\includegraphics[width=\columnwidth]{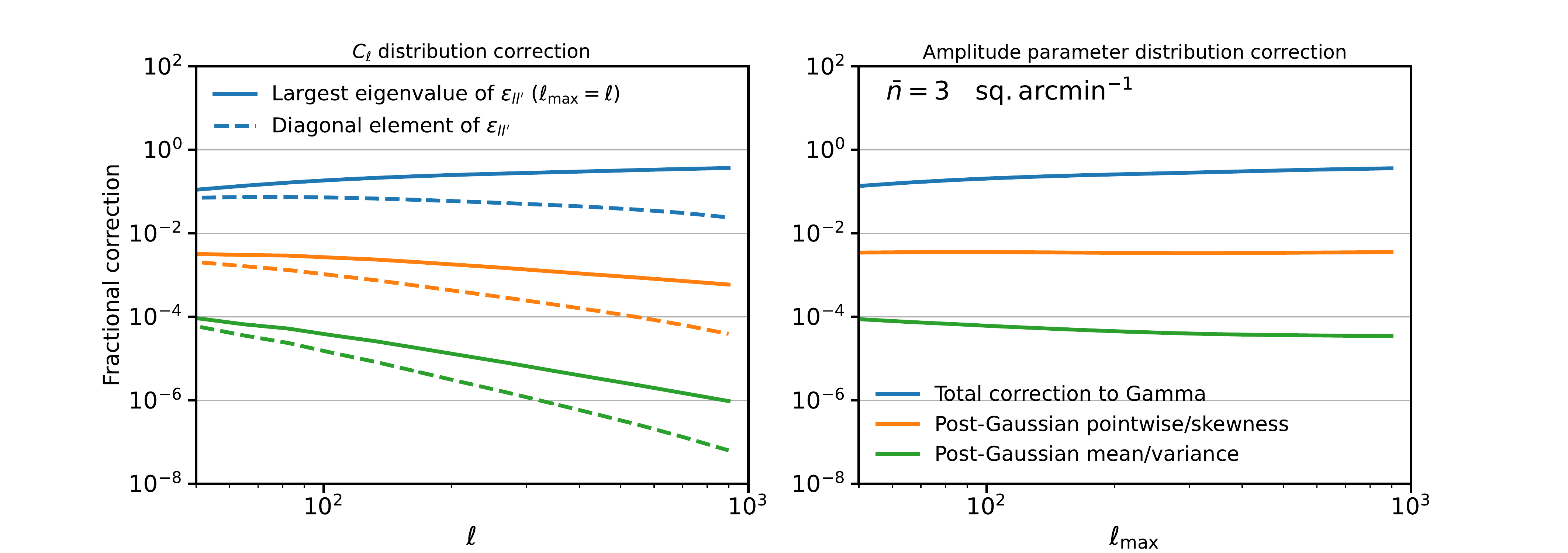}
\caption{Same as Figure~\ref{fig:epsmat_zs6} but with shape noise corresponding to $\bar{n} = 3 \, \mathrm{sq. \, arcmin}^{-1}$.}
\label{fig:epsmat_zs6_n3}
\end{figure}

When shape noise is added the effects of non-Gaussianity are strongly suppressed. Figure~\ref{fig:epsmat_zs6_n3} shows leading-order corrections to the posterior for a model in which $C_l = A(S_{0,l} + N_l)$ when applying shape noise with a galaxy number density of $\bar{n} = 3 \, \mathrm{sq. \, arcmin}^{-1}$, appropriate for a single redshift bin in a Euclid-like survey. Our approximation is now convergent out to much higher $l$, and post-Gaussian corrections are negligible. It thus appears that although the strength of the signal non-Gaussianity may be stronger at low redshift, the relatively higher noise level suppresses non-Gaussianity in the likelihood.

We close this section by noting that these conclusions may need to be modified when considering galaxy clustering at low redshift. There the geometric suppression suffered by lensing is smaller, and hence the Gaussianizing effect of noise may be expected to be relatively weaker. We have also neglected Intrinsic Alignments, which may be expected to play a greater role in cross-bin combinations at low redshift. We advocate that the statistics plotted in Figure~\ref{fig:epsmat} and Figure~\ref{fig:epsmat_zs6} be computed whenever a likelihood approximation needs to be made.

\bibliography{references}

\end{document}